\documentclass{article}
\usepackage{graphicx} 
\usepackage{lineno,hyperref}
\usepackage{amsmath}
\usepackage{amsfonts}
\usepackage{xcolor}
\usepackage{geometry}
\usepackage{textcomp}
\usepackage{comment}
\usepackage{float}

\usepackage[round,sort,comma]{natbib}

\usepackage[table]{xcolor}

\usepackage{capt-of}
\usepackage{booktabs}
\usepackage{varwidth}
\usepackage{amssymb}

\usepackage[ruled,vlined]{algorithm2e}

\usepackage[british]{babel}
\usepackage{hhline}
\usepackage{multirow}
\usepackage{authblk}
\usepackage{soulpos}

\usepackage{soul}
\definecolor{OliveGreen}{rgb}{0,0.6,0}

\usepackage{algorithmic}
\usepackage{array}

\usepackage{geometry}

\usepackage{authblk}
\usepackage{etoolbox} 

\makeatletter
\patchcmd{\@maketitle}{\@author}{\@author\vskip1em}{}{}
\makeatother

\title{\textbf{Shocks Under Control: Taming Transonic Compressible Flow over an RAE2822 Airfoil with Deep Reinforcement Learning}}

\author[1]{Trishit Mondal}
\author[2]{Ricardo Vinuesa}
\author[1]{Ameya D. Jagtap\thanks{Corresponding author: Ameya D. Jagtap (ajagtap@wpi.edu, ameyadjagtap@gmail.com)}}

\affil[1]{\textit{\small{Aerospace Engineering Department, Worcester Polytechnic Institute, Worcester, MA 01609, USA.}}}
\affil[2]{\textit{\small{Department of Aerospace Engineering, University of Michigan, MI 48109, USA.}}}

\date{}
\begin{document}
\maketitle

\begin{abstract}
Active flow control of compressible transonic shock–boundary layer interactions over a two-dimensional RAE2822 airfoil at $R_e = 50{,}000$ is investigated using deep reinforcement learning (DRL). The flow field exhibits highly unsteady dynamics, including complex shock–boundary layer interactions, shock oscillations, and the generation of Kutta waves from the trailing edge. A high-fidelity CFD solver, employing a fifth-order spectral discontinuous Galerkin scheme in space and a strong-stability-preserving Runge–Kutta (5,4) method in time, together with adaptive mesh refinement capability, is used to obtain the accurate flow field.  Synthetic jet actuation is employed to manipulate these unsteady flow features, while the DRL agent autonomously discovers effective control strategies through direct interaction with high-fidelity compressible flow simulations. Unlike conventional approaches relying on reduced-order models or pre-defined actuation schemes, the DRL framework adaptively explores the actuation space to optimize control policies for improved aerodynamic performance. \textit{Three–Three jet} configuration on the upper and lower surfaces of an airfoil is investigated, and distributed environments are employed to enhance exploration strategies. The trained controllers effectively mitigate shock-induced separation, suppress unsteady oscillations, and manipulate aerodynamic forces under transonic conditions.
In the first set of experiments, aimed at both drag reduction and lift enhancement, the DRL-based control reduces the average drag coefficient by 13.78\% and increases lift by 131.18\%, thereby improving the lift-to-drag ratio by 121.52\%, which underscores its potential for managing complex flow dynamics.
In the second set, targeting drag reduction while maintaining lift, the DRL-based control achieves a 25.62\% reduction in drag and a substantial 196.30\% increase in lift, accompanied by markedly diminished oscillations. In this case, the lift-to-drag ratio improves by 220.26\%. These results highlights the intrinsic challenge of reducing drag without compromising lift in transonic flow, where strong shock–boundary-layer interactions couple drag reduction to variations in pressure distribution and lift. 
\end{abstract}

\vspace{0.2cm}

 \begin{small}Keywords: \textit{Deep Reinforcement Learning}; \textit{Active Flow Control}; \textit{Compressible Transonic Flow}; \textit{Shock-Boundary Layer Interactions}.
\end{small}

\section{Introduction}
Most commercial passenger flights operate in the transonic regime, where Mach numbers are typically between 0.8 and 1.0. This makes understanding and controlling transonic aerodynamics critically important for improving efficiency, safety, and fuel economy in aviation. Shock–boundary layer interactions (SBLI) are central to the dynamics of compressible transonic flows and remain a longstanding challenge in fluid mechanics. When a shock impinges on a boundary layer, it induces separation, unsteady oscillations, and large pressure fluctuations that degrade aerodynamic performance and structural integrity. These interactions govern drag rise, buffet onset, and stability margins in transonic airfoils, wings, and intakes, directly influencing the efficiency and safety of high-speed vehicles. Despite decades of research, effective control strategies for mitigating SBLI remain elusive, as the problem is inherently nonlinear, multi-scale, and sensitive to flow conditions. Synthetic jets have emerged as a promising actuation mechanism for active flow control in such regimes. By introducing zero-net-mass-flux disturbances, they enhance boundary layer momentum exchange without requiring complex plumbing or additional mass injection. However, identifying effective actuation strategies remains a significant challenge, since the optimal frequency, amplitude, and spatial distribution of synthetic jet actuation depend strongly on the unsteady shock–boundary layer dynamics.
In this work, we perform active flow control over a two-dimensional airfoil by leveraging deep reinforcement learning (DRL) to discover control strategies for transonic SBLI using synthetic jets. Unlike traditional approaches that rely on reduced-order models or pre-defined actuation schemes, DRL agents interact directly with high-fidelity simulations, adaptively explore the actuation space, and autonomously learn strategies that optimize flow control objectives.

\subsection{Related Work}
The development of DRL for active flow control (AFC) using synthetic jets has evolved steadily over the last two decades, moving from fundamental actuator physics toward advanced intelligent control of complex flows. The concept of synthetic jets, also called zero-net-mass-flux jets, became established in the early 2000s when researchers clarified the physical mechanism of jet formation and established synthetic jets as an effective actuator for AFC applications. A synthetic jet consists of a cavity and an oscillating diaphragm that alternately blows and ingests fluid through an opening, which produces vortex shedding without net mass addition. Early work, such as \cite{holman2005formation} established the physical foundations of how these actuators generate coherent vortical structures. Numerous experiments and simulations demonstrated that synthetic jets could delay or suppress flow separation when appropriately tuned on airfoils and bluff bodies. Since then, researchers have investigated how jet orientation, cavity geometry, and actuation waveform influence vortex generation and interaction with the external flow.

\cite{choi2008control} provided an early foundation by reviewing active and passive control strategies for bluff-body flows, emphasizing wake manipulation and efficiency across Reynolds-number regimes. More than a decade later, \cite{ecke20172d} analyzed dimensional transitions between two- and three-dimensional turbulence, offering physical insights relevant to flow-control design. Later, \cite{pivot2017continuous} primarily demonstrated the potential of a continuous RL framework for closed-loop drag reduction in cylinder wake flows, where an implicit model was learned from sparse sensor data. \cite{reddy2018glider} then advanced RL concepts to autonomous soaring, where an RL-trained glider learned to navigate atmospheric thermals using real-world sensory cues such as vertical accelerations and roll torques. The introduction of DRL into AFC was marked by \cite{rabault2019artificial}, who integrated modern DRL techniques with traditional AFC concepts, replacing pre-programmed harmonic forcing with self-learned, closed-loop control. Extending DRL beyond steady aerodynamic flows, \cite{novati2019controlled} used reinforcement learning to learn gliding and perching maneuvers for elliptical bodies, discovering energy- and time-optimal trajectories without explicit physical modeling. Experimental demonstrations followed, such as \cite{shimomura2020closed}, who applied a deep Q-network for closed-loop separation control over a NACA0015 airfoil using plasma actuators, achieving sustained flow attachment superior to open-loop schemes. \cite{fan2020reinforcement} validated RL-based AFC experimentally for bluff bodies, optimizing cylinder rotation to reduce turbulent drag.

Progress in 2021 included \cite{ren2021applying}, who extended DRL-based AFC to weakly turbulent cylinder wakes at $Re=1000$, showing effective drag reduction, and \cite{qin2021application}, who introduced a data-driven reward function using dynamic mode decomposition, enabling DRL agents to extract global flow features for drag reduction and enhanced recirculation. \cite{mei2021active} applied DRL to actively control jet flow over a circular cylinder to enhance vortex-induced vibration for energy harvesting. In the same year, \cite{zheng2021active} compared active learning and DRL strategies for suppressing vortex-induced vibration, finding the soft actor-critic DRL approach most effective. Beyond aerodynamic control, \cite{xie2021sloshing} applied DRL with behavior cloning to suppress liquid sloshing in tanks, achieving over 80\% oscillation reduction. \cite{mandralis2021learning} leveraged RL to reproduce energy-efficient escape responses in larval fish, identifying biologically realistic movement strategies.

In 2022, the range of DRL applications expanded rapidly. \cite{wang2022deep} used the Proximal Policy Optimization (PPO) algorithm to control disturbed flow over a NACA0012 airfoil, improving both drag and lift while optimizing actuation energy. Reviews by \cite{vinuesa2022flow} and \cite{chen2022review} summarized progress in data-driven flow control and emerging AI-driven methods for vortex suppression. Complementarily, \cite{konishi2022fluid} employed RL for enhancing passive-scalar mixing, where agents learned to exploit stagnation zones for efficient mixing. Focusing on transferability, \cite{wang2022accelerating} introduced transfer learning for PPO-based flow-control agents, significantly reducing training cost, while \cite{kubo2022efficient} proposed DRL control robust to partial observability. In the same year, \cite{yu2022deep} explored collective swimming coordination via DRL, showing cooperative energy-efficient strategies, and \cite{amico2022deep} experimentally demonstrated DRL control of 3-D bluff-body wakes at $Re \approx 10^5$.

\cite{dobakhti2023active} expanded the complexity of DRL-controlled flow systems by combining cylinder rotation with multiple synthetic jets, achieving up to 50\% drag reduction and highlighting the role of optimal sensor placement. \cite{jiang2023reinforcement} investigated rotational oscillation control of circular cylinders via PPO, reporting 21\% drag reduction. Similarly, \cite{he2023policy} demonstrated transfer of PPO-trained 2-D DRL agents to 3-D cylinder wakes, validating cross-dimensional generalization. In parallel, \cite{paris2023reinforcement} proposed a DRL-based actuator-selection framework with adaptive sparsification to minimize actuator use without degrading control performance.  \cite{guastoni2023deep} give the first application of DRL to control in a turbulent channel.
\cite{wang2024dynamic} introduced a dynamic-feature–based DRL framework coupling the Soft Actor-Critic algorithm with pressure-based dynamic feature extraction for cylinder wake control, enabling effective drag reduction from sparse wall-pressure data. \cite{jia2024effect} systematically analyzed geometric influences of synthetic jets (placement and slot width) on DRL performance over square cylinders, highlighting the need for co-optimization between actuator geometry and learned policy.

By 2025, DRL-based flow control had reached high-Reynolds-number and fully turbulent regimes. \cite{chen2025active} demonstrated DRL-based control of turbulent bluff-body flows at $Re=2.74\times10^5$, achieving 29\% drag reduction and robustness across regimes. \cite{font2025deep} controlled turbulent separation bubbles via synthetic jets, outperforming the best open-loop forcing and proving DRL policies transferable from coarse to fine CFD grids. Extending into three dimensions, \cite{suarez2025flow} implemented a multi-agent RL framework for spanwise-distributed synthetic jets around cylinders, discovering coordinated actuation strategies with reduced mass flux. \cite{yan2025deep} developed a mutual-information-based SAC method transferring control policies from 2-D to 3-D bluff-body flows, drastically reducing training cost. Further, \cite{zhou2025reinforcement} applied DRL to turbulent channel flows (friction Reynolds number $Re_\tau=1000$), achieving 30\% drag reduction through suppression of near-wall streaks. To this end, \cite{montala2025deep} demonstrated DRL control of separated flows over a NACA0012 wing at $Re=1000$ and $20^\circ$ angle of attack, where the PPO-trained agent autonomously modulated leading-edge vortex shedding for lift enhancement and separation mitigation.

In compressible aerodynamics, recent studies have demonstrated the promise of DRL for active flow control. ~\cite{ren2024adaptive} employed DRL to suppress transonic buffet and buffeting phenomena, enabling the agent to mitigate unsteady aerodynamic loads and reduce pitching vibrations in coupled fluid-structure interaction systems without the need for explicit dynamic models. Similarly, ~\cite{tao2025control} achieved DRL-based control of hypersonic shock-boundary-layer interactions (for neural flow) using microjet actuation, effectively reducing separation extent and training cost through a model-informed strategy. In both cases, the authors employed second-order fluid flow solvers. Recently, \cite{zong2025closed} developed an FPGA-based experimental DRL control framework that runs at 1-10~kHz, that is 100 times faster than CPU-based systems, and demonstrated its effectiveness on supersonic backward-facing step flow, achieving a 21.2\% increase in shear-layer mixing with only 10 seconds of training.

\subsection{Contribution of this work}
This study reports the first application of DRL to the active control of transonic shock–boundary-layer interaction using a high-fidelity CFD solver. Integrating such a solver with DRL-based control poses significant challenges, owing to the solver’s computational complexity and the need for stable interaction. Following are the contributions of the present work:
\begin{itemize}
    \item A fully coupled DRL framework is developed and integrated with high-fidelity compressible Navier–Stokes simulations of the two-dimensional RAE2822 airfoil at $R_e = 50{,}000$. The solver employs a fifth-order spectral discontinuous Galerkin discretisation, an SSPRK54 time-stepping scheme, and adaptive mesh refinement, enabling accurate resolution of unsteady phenomena such as shock oscillations, Kutta wave generation, and boundary-layer separation.
    \item The \textit{Three–Three} synthetic-jet configuration, positioned on the upper and lower surfaces of the airfoil, is investigated. A distributed training environment is employed to enhance exploration efficiency.
   \item The first objective is to reduce drag and enhance lift of the RAE2822 airfoil. A DRL agent autonomously learns effective synthetic-jet actuation strategies through direct interaction with the flow field, without reliance on reduced-order models, linearized assumptions, or predefined control laws. The trained control mitigates shock-induced separation, suppresses unsteady shock oscillations, and improves aerodynamic performance under transonic conditions.
   \item The second objective is to reduce drag while maintaining lift near the baseline value, a condition particularly relevant to cruise flight regimes typical of commercial aircraft operation.
\item We further compare the on-policy \textit{Proximal Policy Optimization} (\cite{schulman2017proximal}) and the off-policy \textit{Twin Delayed Deep Deterministic Policy Gradient} (\cite{fujimoto2018addressing}) methods.

\end{itemize}
This paper is organized as follows. In Section 2, we present the governing equations and problem formulation, including the computational domain and boundary conditions for compressible transonic flow over the RAE2822 airfoil. Section 3 details the high-fidelity CFD solver and presents the associated convergence study. Section~4 details the deep reinforcement learning methodology, followed by Section~5, which presents results for various configurations, including distributed environments and using both on- and off-policy DRL algorithms. To this end, we summarize our findings in Section 6.

\section{Governing Equations and Problem Formulation}
The compressible Navier–Stokes equations are used to simulate transonic flow over the two-dimensional RAE2822 airfoil and can be expressed in conservative form as follows:
\begin{equation}\label{NSE}
\frac{\partial \mathbf{U}}{\partial t} + \frac{\partial \mathbf{F}_{\text{inv}}}{\partial x} + \frac{\partial \mathbf{G}_{\text{inv}}}{\partial y} = \frac{\partial \mathbf{F}_v}{\partial x} + \frac{\partial \mathbf{G}_v}{\partial y},  ~~ t \in \mathbb{R}^+ ~\text{and}~ (x,y) \in \Omega \subset \mathbb{R}^2,
\end{equation}
where the conserved variables vector $\mathbf{U} $ and the inviscid flux vectors 
$\mathbf{F_{\text{inv}}}, \mathbf{G_{\text{inv}}}$ are given as
\begin{equation}
  \mathbf{U} = \begin{bmatrix}
    \rho \\
    \rho u\\
    \rho v\\
    \rho E
\end{bmatrix} ;
\quad
\mathbf{F_{\text{inv}}} =  \begin{bmatrix}
    \rho u\\
    \rho u^2 + p\\
    \rho uv\\
    u(\rho E + p)
\end{bmatrix};
\quad
\mathbf{G_{\text{inv}}} = \begin{bmatrix}
    \rho v \\
    \rho uv\\
    \rho v^2 + p\\
    v(\rho E + p)
\end{bmatrix}.
\end{equation}

The viscous flux vectors $\mathbf{F}_v,\mathbf{G}_v$ are given as
\begin{equation}
\mathbf{F}_v = \begin{bmatrix}
    0\\
    \tau_{xx}\\
    \tau_{xy}\\
    u\tau_{xx} + v\tau_{xy} + q_x 
\end{bmatrix};
\quad
\mathbf{G}_v = \begin{bmatrix}
    0\\
    \tau_{xy}\\
    \tau_{yy}\\
    u\tau_{xy} + v\tau_{yy} + q_y
\end{bmatrix}.
\end{equation}
Here, $\rho$ represents the density, $\mathbf{u} = \{u,v\}$ is the velocity vector, $p$ is the pressure, and $E = \frac{p}{\rho(\gamma-1)} + \frac{1}{2} \|\mathbf{u}\|_2^2$ is the total 
energy per unit mass. The viscous stress tensor $\boldsymbol{\tau}$ characterizes 
momentum diffusion due to viscosity $\mu$, while $ q_x = -k\frac{\partial T}{\partial x}, \quad q_y = -k\frac{\partial T}{\partial y}
$ represents the heat flux 
with thermal conductivity $k$. 
The ideal gas equation of state $p=\rho RT $ involves the temperature $T$, the specific gas constant $R$, specific heat at constant volume $c_v$, specific heat at constant pressure $c_p$, and their ratio $\gamma = c_p/c_v$. The viscous stress tensor components are given by
\begin{align*}
\tau_{xx} &= \mu\left(2\frac{\partial u}{\partial x} - \frac{2}{3}\nabla \cdot \mathbf{u}\right), \quad
\tau_{yy} = \mu\left(2\frac{\partial v}{\partial y} - \frac{2}{3}\nabla \cdot \mathbf{u}\right) \\
\tau_{xy} &= \mu\left(\frac{\partial u}{\partial y} + \frac{\partial v}{\partial x}\right).
\end{align*}

\begin{figure}[h]
    \centering
    \includegraphics[scale=0.7, clip=true]{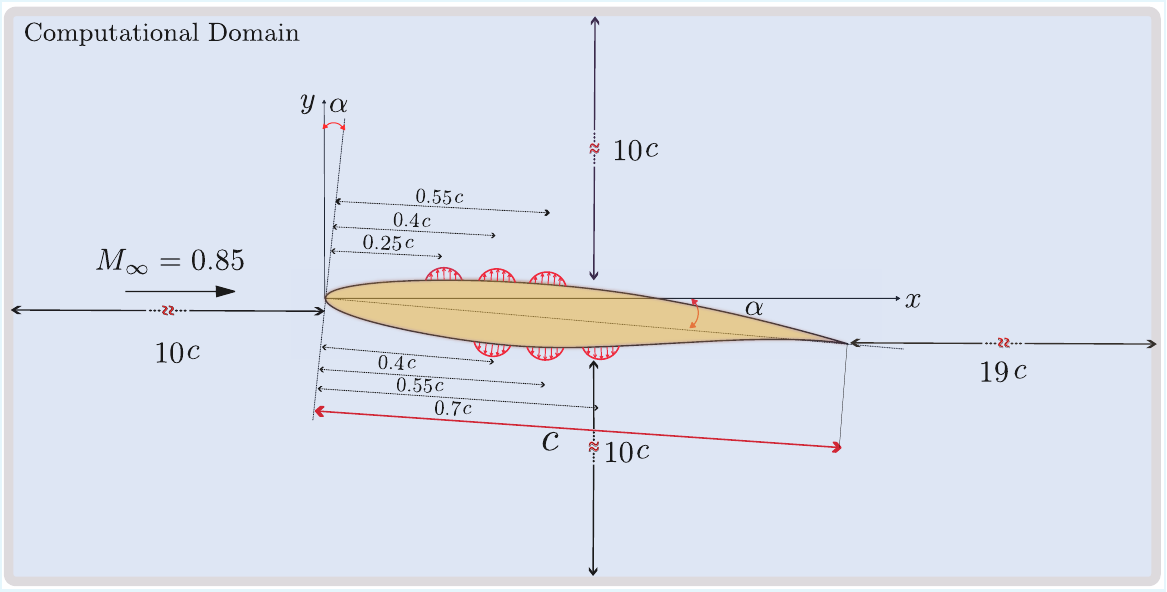}
    \caption{Compressible transonic flow over the RAE2822 2D airfoil with synthetic jet actuation. Three synthetic jets are depicted on both the upper and lower surfaces of the airfoil.}
    \label{fig: jets} \end{figure}

We consider the problem of transonic flow over a two-dimensional RAE2822 airfoil, where shock–boundary layer interactions lead to flow separation, unsteadiness, and a loss of aerodynamic performance. Controlling these nonlinear and unsteady interactions presents a significant challenge.
Figure \ref{fig: jets} shows the schematic representation of a computation domain for RAE2822 airfoil with 6 jets (3 on top surface and 3 on bottom). Due to the transonic nature of the flow, the computational domain was extended to minimize spurious reflections from the boundaries. At the far-field boundaries, where waves may enter or exit the computational domain, the convective fluxes are treated using a characteristic-based formulation: incoming waves are specified from the freestream, while outgoing waves are extrapolated from the interior, thereby implementing Riemann-invariant boundary conditions. Supersonic inflow and outflow regions are treated by fully prescribing or extrapolating the flow variables, respectively. On the airfoil surface, the viscous boundary layer is resolved by enforcing a no-slip condition on the velocity and adiabatic condition on the temperature, with the pressure and density evolving according to the governing equations. This combination of characteristic far-field treatment and physically consistent wall conditions ensures that both the hyperbolic and parabolic aspects of the compressible Navier–Stokes equations are accurately represented.

\section{High-Fidelity CFD Solver}
We employed the fifth-order spectral DG method (\cite{hesthaven2008nodal}) for space discretization, solving governing compressible Navier-Stokes equations. In the DG formulation, the computational domain is discretized into non-overlapping elements. 
Within each element, the solution is approximated using high-order polynomial basis functions. 
The weak spectral DG formulation for compressible Navier-Stokes equations is given in Appendix \ref{appA}. The semi-discrete spectral DG formulation results in a system of ordinary differential equations in time. This system is integrated in time using the SSPRK54 (\cite{spiteri2002new}). Appendix \ref{appB} gives detail description of SSPRK method.

To efficiently capture flow features with varying scales, such as shocks, boundary layers, and wakes, we employ adaptive mesh refinement (AMR) using the \texttt{P4estMesh} library \cite{burstedde2011p4est}. The computational domain is represented as a parallel forest-of-quadtrees, where each quadrilateral element can be recursively subdivided to create multiple refinement levels. Mesh refinement is guided by a L\"ohner indicator (\cite{Lohner1987}) based on density gradients,
\[
\eta_i = \frac{|\nabla^2 \rho_i|}{|\nabla \rho_i| + \epsilon},
\]
which effectively identifies shocks and steep gradients. The DG method handles non-conforming interfaces via numerical fluxes, ensuring conservation and stability, while \texttt{P4estMesh} manages neighbor connectivity and load balancing for scalable parallel computations. This concentrates resolution near the airfoil and shocks while keeping coarser grids elsewhere, improving efficiency without sacrificing accuracy. 

The computational mesh used in all test cases is an unstructured quadrilateral grid comprising 1,800 elements in the initial configuration. The number of elements increases dynamically during the simulations due to AMR.

The higher-order spectral DG scheme, combined with SSPRK54 time integration and adaptive mesh refinement, offers several key advantages, including robust treatment of high-gradient features such as shocks, strict local conservation, and a natural framework for dynamic mesh adaptation. When applied to the RAE2822 airfoil, this approach accurately resolves strong pressure gradients near shock–boundary interactions and captures detailed wake dynamics downstream. Compared with conventional methods, the spectral DG formulation provides sharper shock resolution, reduced numerical dissipation in the wake, and improved predictive accuracy for critical aerodynamic coefficients, including lift and drag, $C_l$ and $C_d$.

\subsection{Convergence Study (No Jets)}
To assess the convergence of the higher-order scheme, we simulated transonic flow over the RAE2822 airfoil using polynomial orders $p = 5$, 6, and 7 on an initial mesh generated with \texttt{Gmsh} \cite{geuzaine2009gmsh}.
The dimensionless flow parameters correspond to transonic conditions are $M_\infty = 0.85, \quad \alpha = 2^\circ, \quad \text{Re} = 5.0 \times 10^4, \quad 
p_\infty = 1.0, \quad T_\infty = 1.0, \quad \gamma = 1.4, \quad R = 287.87.$
The reference quantities are computed from the free-stream conditions: the density from the ideal gas law, $\rho_\infty = p_\infty/(R T_\infty) \approx 3.474\times10^{-3}$, the velocity, $u_\infty = 17.07$, and the dynamic viscosity from the Reynolds number $R_e = 50,000$, $\mu \approx 1.19\times10^{-6}$, here $c = 1$ is the chord length.

\begin{figure}[htpb]
    \centering
    \includegraphics[scale=0.74, clip=true]{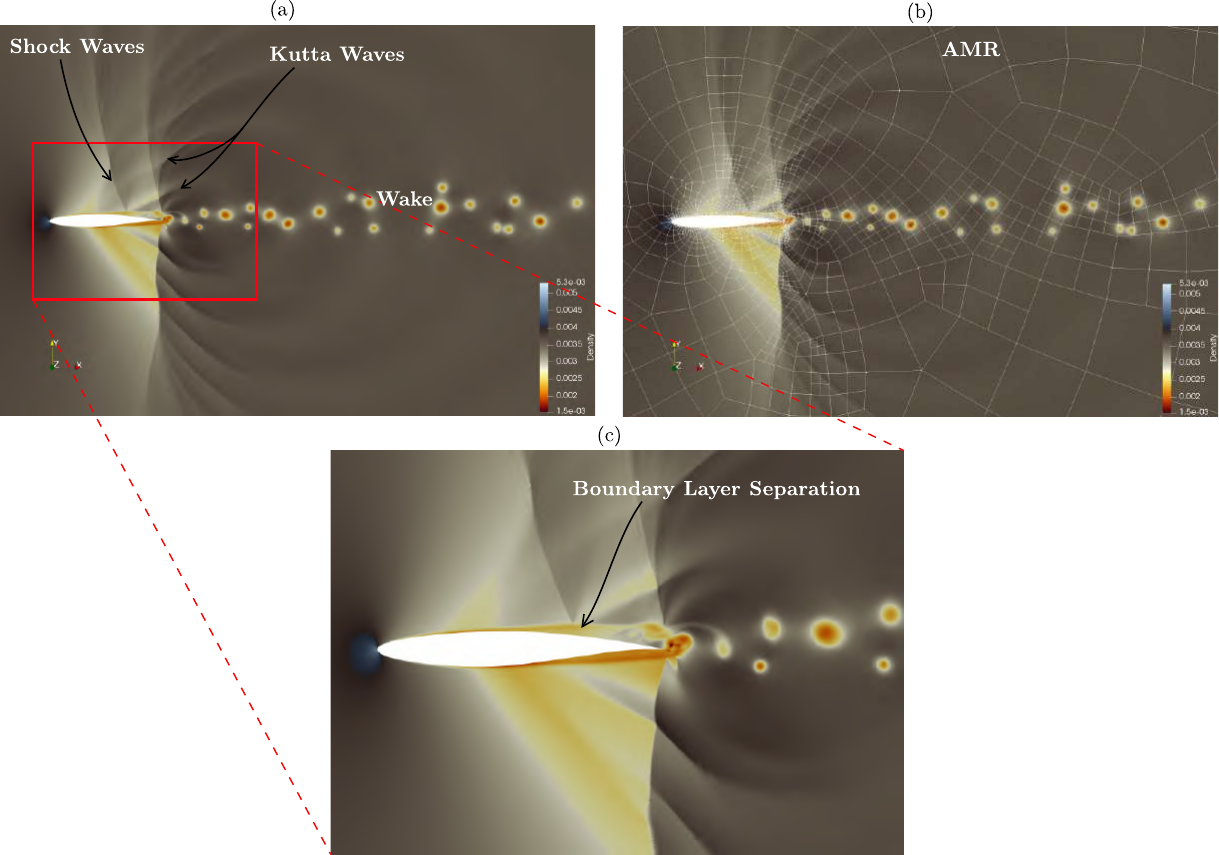}
    \caption{Density contours illustrating flow features around the airfoil.
(a) Density contours showing Kutta waves interacting with shock waves and the wake region. (b) Density contours computed with adaptive mesh refinement, highlighting enhanced resolution in critical regions. (c) Zoomed view near the airfoil surface, revealing boundary layer separation.} 
    \label{fig:jets}
\end{figure}
\begin{figure}[htpb]
    \centering
    \includegraphics[scale=0.33, clip=true]{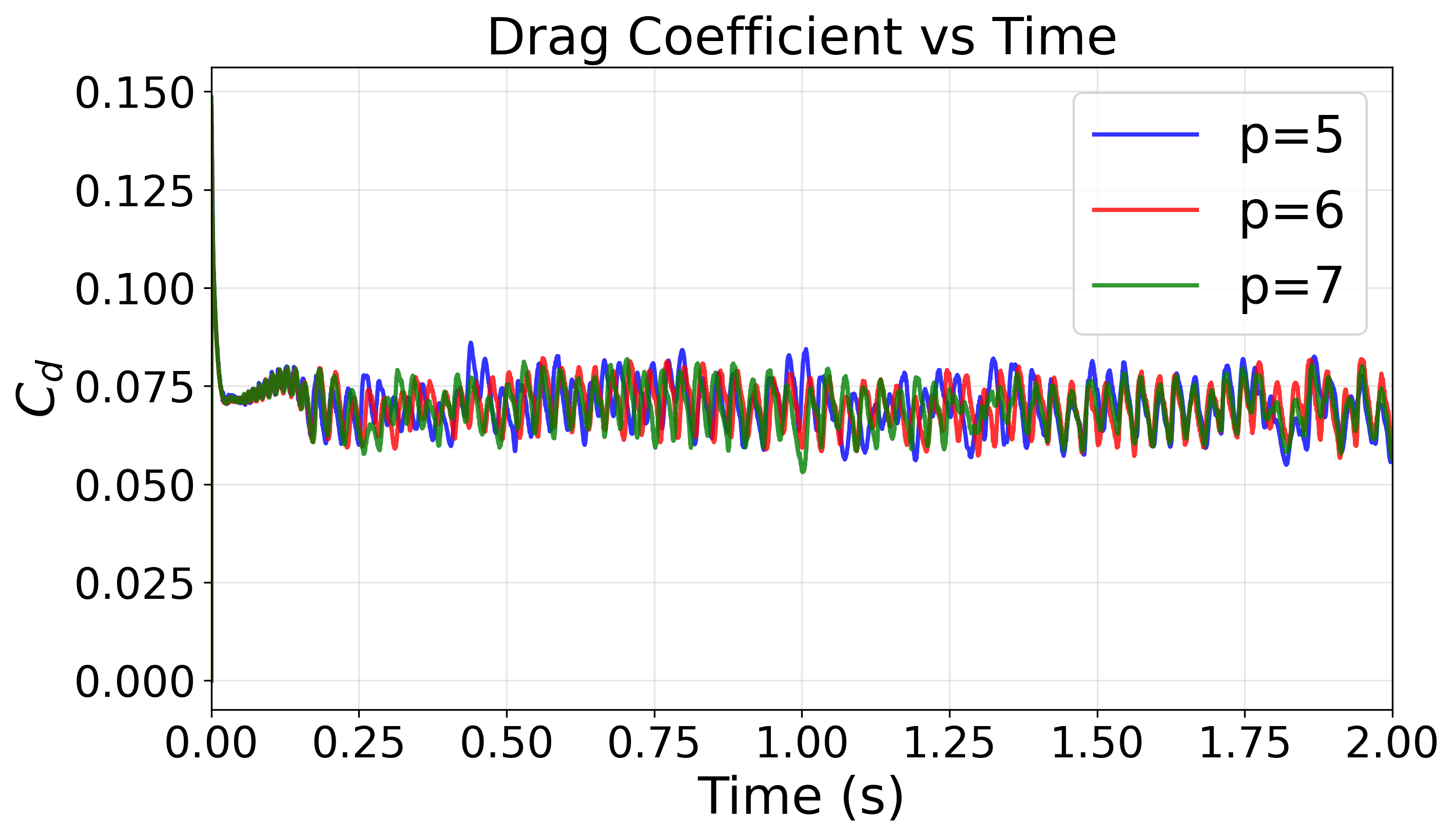}
    \includegraphics[scale=0.33, clip=true]{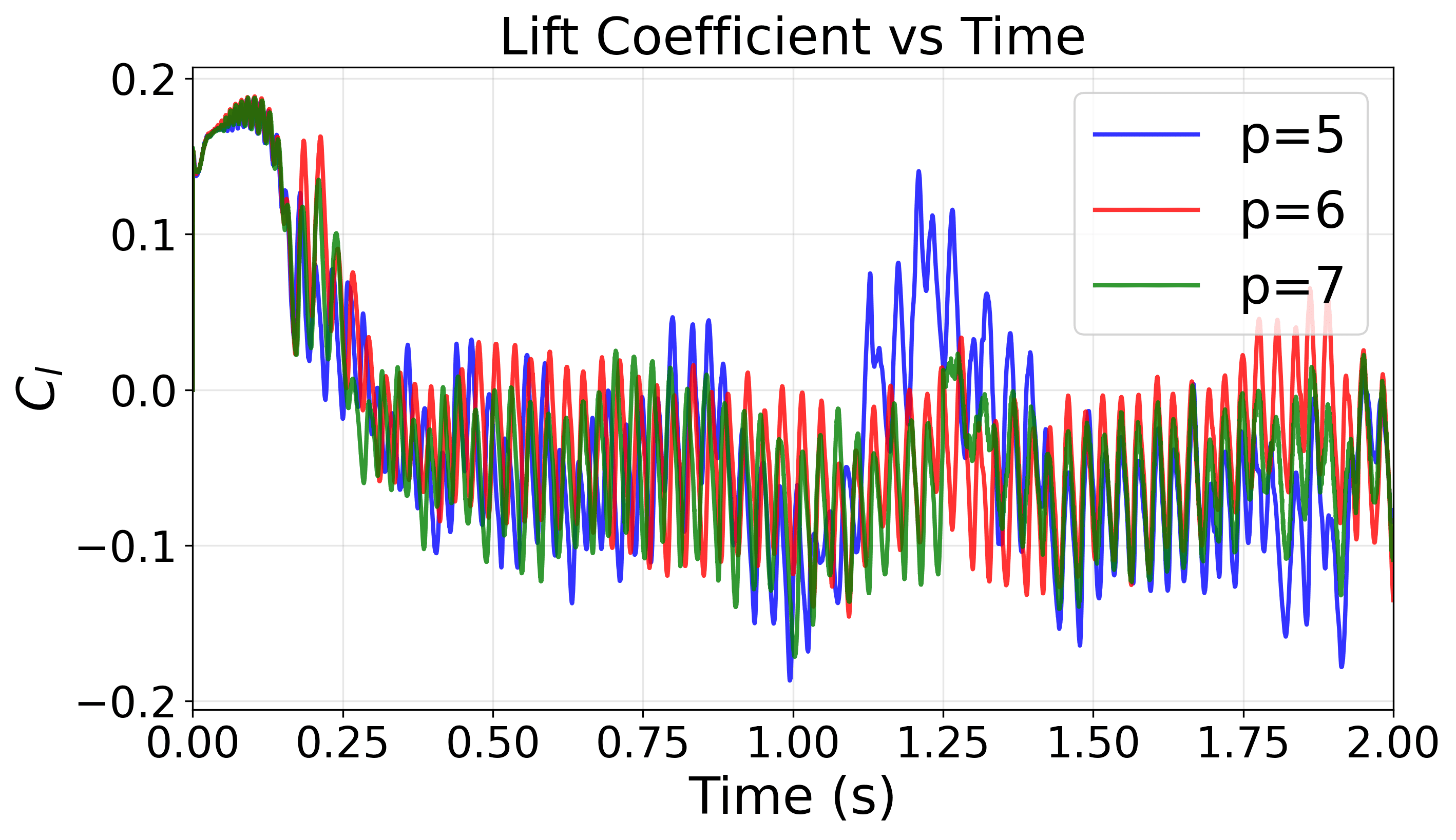}
    \caption{Drag and Lift coefficients for 5th, 6th, and 7th order polynomials.}
    \label{fig:comparison_cl_cd_poly}
\end{figure}

Density contours in figure~\ref{fig:jets} illustrates flow features around the airfoil. The Kutta waves generated at the airfoil trailing edge propagate and interact with shock waves on both the upper and lower surfaces, producing complex unsteady flow patterns. The flow near the airfoil surfaces exhibits pronounced SBLI, where the impinging shocks induce boundary layer thickening and local flow deceleration. In regions of strong adverse pressure gradient, this can trigger boundary layer separation, forming recirculation zones and increasing wake turbulence. The high-resolution AMR mesh captures Kutta waves, the shock structure, and the boundary layer response, resolving the separated flow regions and shear-layer dynamics. 
The lift and drag coefficients are computed from the surface pressure distribution on the airfoil:
\begin{align}
C_l &= \frac{1}{\frac{1}{2} \rho_\infty u_\infty^2 c} \int_{\text{S}} p_S \, \hat{\mathbf{n}} \cdot \mathbf{e}_l \, dS, \\
C_d &= \frac{1}{\frac{1}{2} \rho_\infty u_\infty^2 c} \int_{\text{S}} p_S \, \hat{\mathbf{n}} \cdot \mathbf{e}_d \, dS,
\end{align}
where $p_S$ is the pressure on the airfoil surface, $\hat{\mathbf{n}}$ is the outward normal, and the lift and drag directions are 
\[
\mathbf{e}_l = \begin{pmatrix}-\sin\alpha \\ \cos\alpha\end{pmatrix}, \qquad 
\mathbf{e}_d = \begin{pmatrix}\cos\alpha \\ \sin\alpha\end{pmatrix}, \quad \alpha = 2^\circ.
\]
Figure~\ref{fig:comparison_cl_cd_poly} compares the lift and drag coefficients, $C_l$ and $C_d$, for polynomial orders $p = 5, 6,$ and $7$. As expected, both coefficients exhibit oscillatory behaviour. Moreover, the drag coefficient remains largely unchanged across all polynomial orders, whereas the lift coefficient exhibits slight variations, reflecting the increased sensitivity of lift to the solution’s local resolution. 

To capture turbulence without explicit subgrid-scale models, the solver employs an implicit Large Eddy Simulation (iLES) approach. In this framework, the high-order discontinuous Galerkin discretization, combined with conservative adaptive mesh refinement, provides sufficient numerical dissipation to emulate subgrid-scale energy transfer. This enables the direct resolution of large-scale turbulent structures, while the smaller, unresolved scales are effectively represented by the solver’s intrinsic numerical dissipation, demonstrating that iLES is adequate for accurate turbulence modeling in this high-fidelity framework. To further assess the solver’s capability to capture turbulent dynamics, the kinetic energy spectra are computed in the wake region, which is rich in energetic scales and characterized by vortex shedding and strong shear interactions. Analyzing the spectra in this region quantifies the energy distribution across scales, confirms that large-scale vortices are well resolved, and illustrates that the solver’s inherent dissipation accurately represents the smaller, unresolved scales without requiring explicit subgrid-scale modeling.

\begin{figure}[htpb]
    \centering
    \includegraphics[scale=0.24, clip=true]{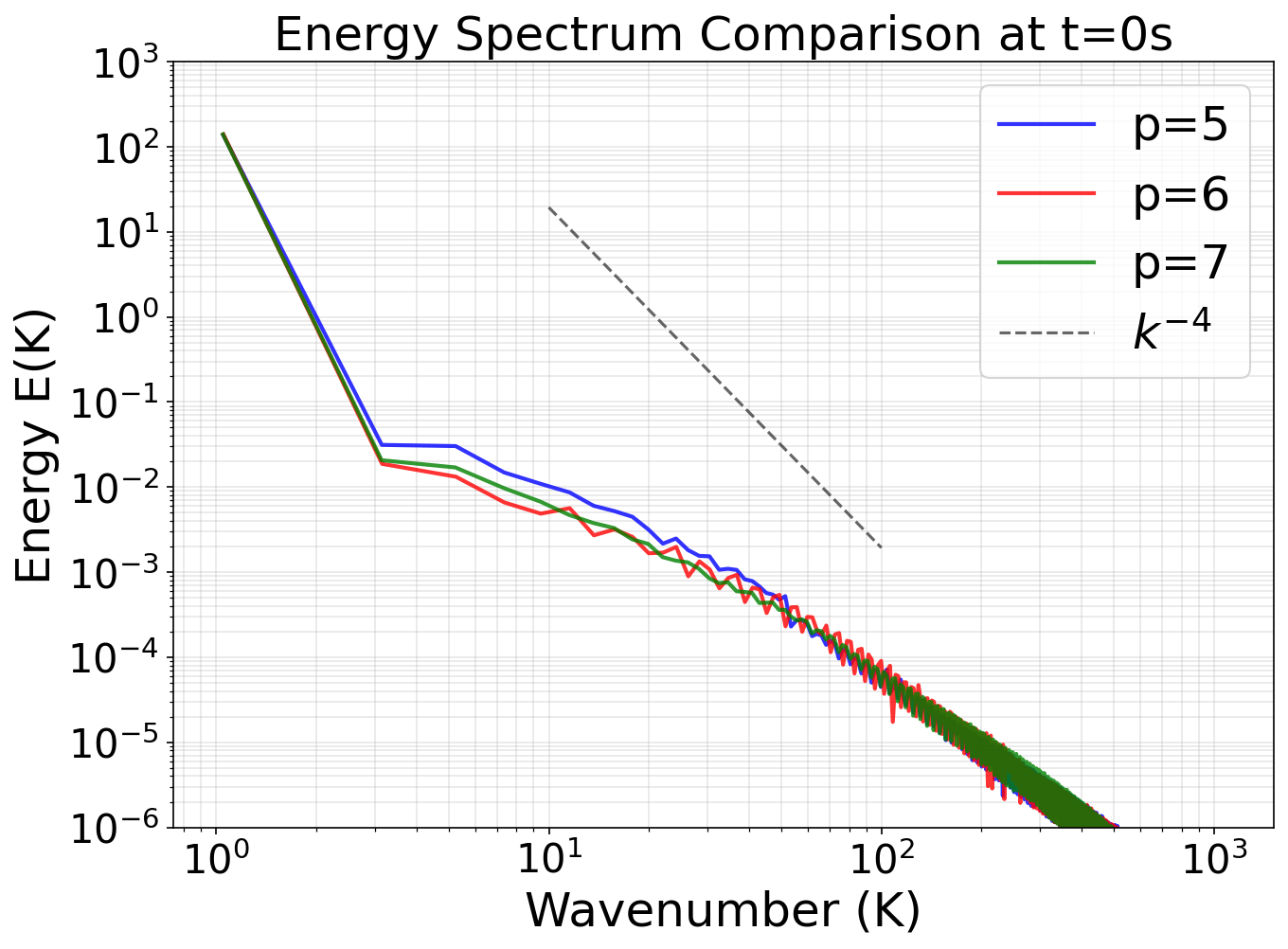}
    \includegraphics[scale=0.24, clip=true]{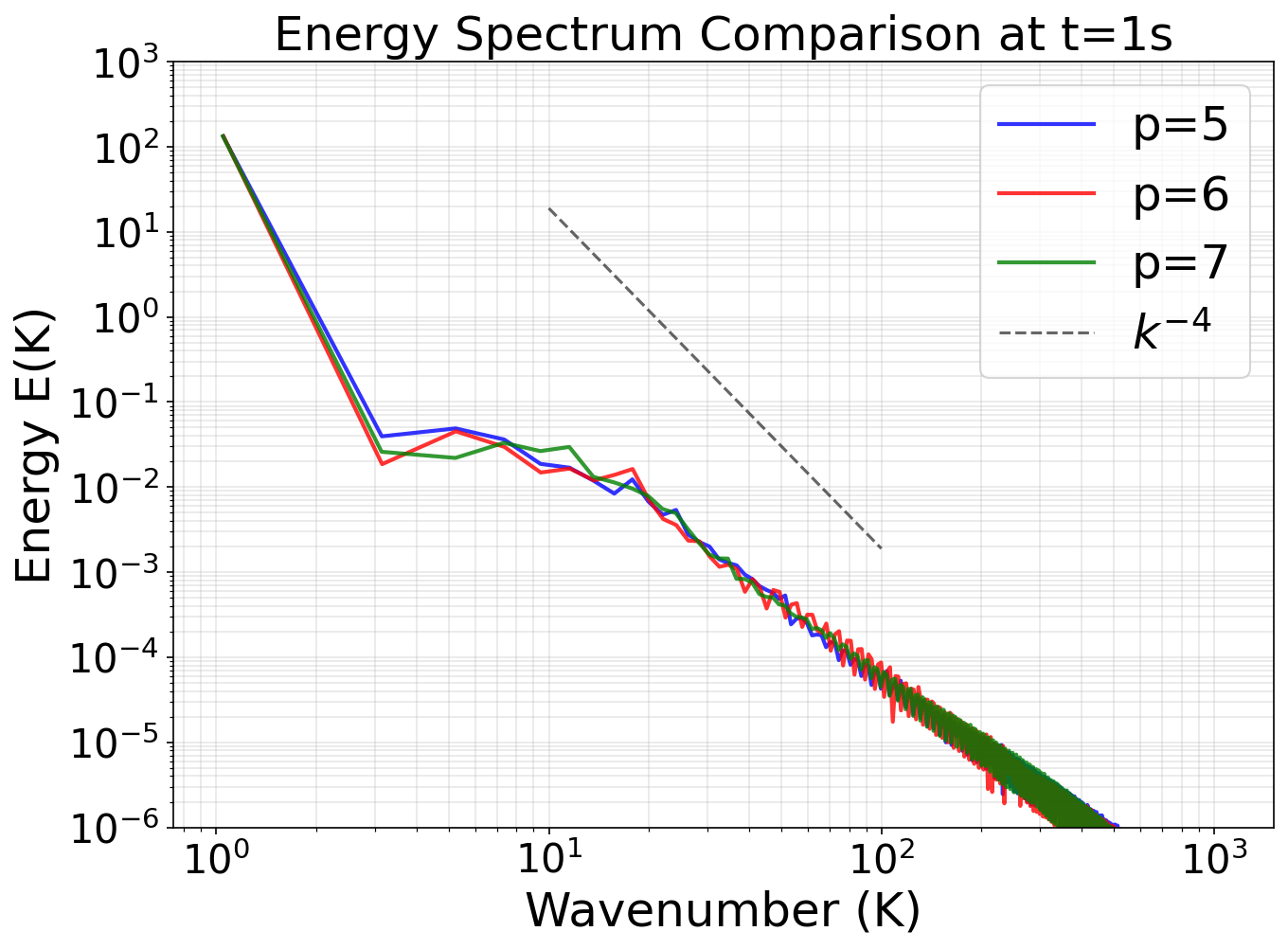}
        \includegraphics[scale=0.24, clip=true]{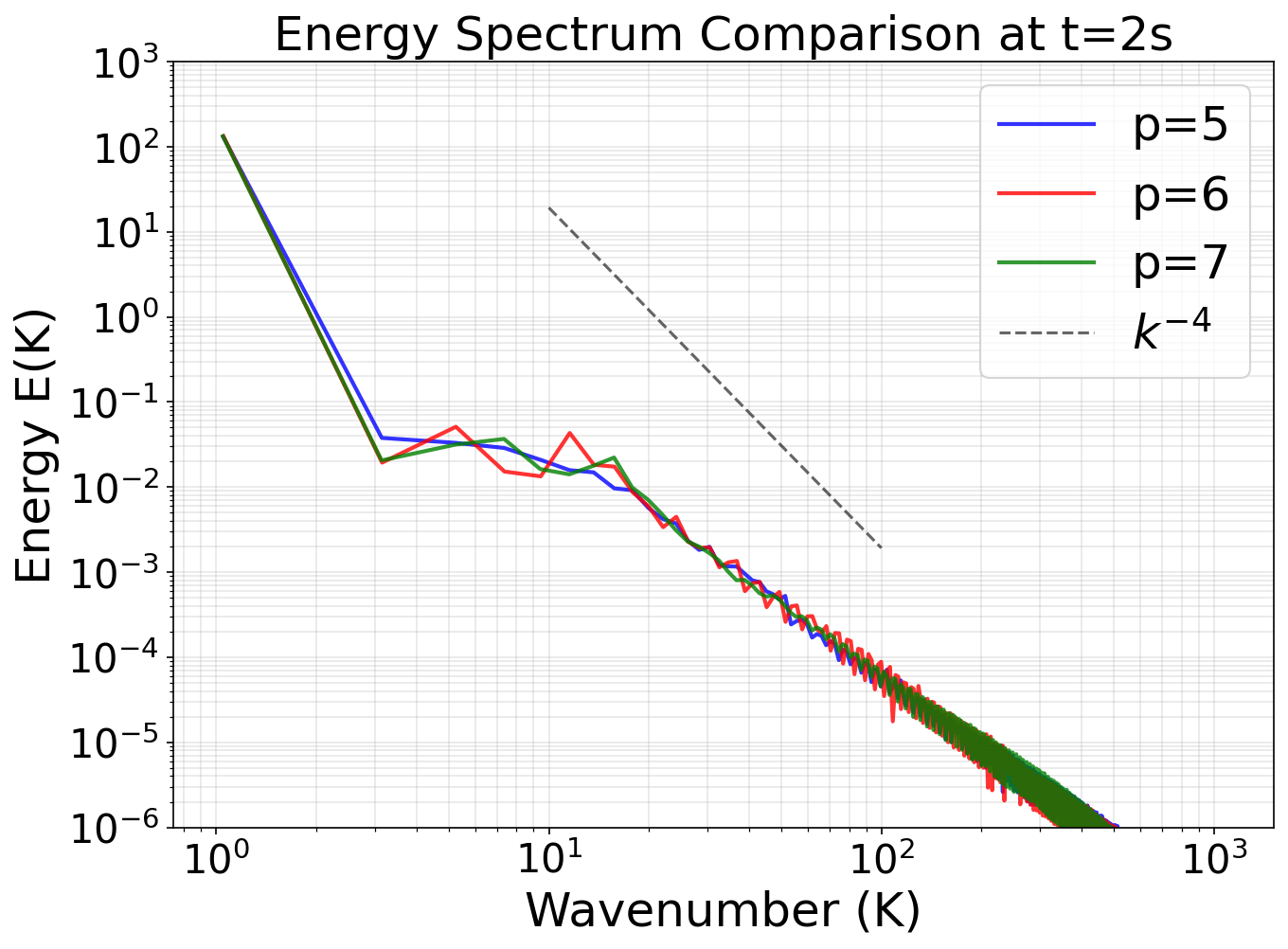}
    \caption{The kinetic energy spectra for polynomial orders $p = 5, 6,$ and $7$ at $t = 0$, $1$, and $2$ seconds, illustrating the convergence of the resolved scales. }
    \label{fig:comparison_k_spectra}
\end{figure}
The kinetic energy spectra are computed from a two-dimensional Fourier transform of the velocity components $u$ and $v$: $E_{2D}(\mathbf{k}) = \frac{1}{2}\left(|\hat{u}(\mathbf{k})|^2 + |\hat{v}(\mathbf{k})|^2\right)$,
where $\hat{u}$ and $\hat{v}$ denote the Fourier-transformed velocities, and $\mathbf{k} = (k_x, k_y)$. The corresponding one-dimensional spectrum, $E(K)$, is obtained by summing the energy within a shell of radius $K = |\mathbf{k}|$:
$E(K) = \sum_{K - \Delta K/2 < |\mathbf{k}| \leq K + \Delta K/2} E_{2D}(\mathbf{k})$.
This shell-averaging reduces the two-dimensional distribution to a function of wavenumber magnitude. Figure~\ref{fig:comparison_k_spectra} shows the resulting spectra for polynomial orders $p = 5, 6,$ and $7$ at $t=0,1,$ and 2 seconds, demonstrating convergence of the resolved scales.

\section{Deep Reinforcement-Learning Setup}
We cast the transonic flow control problem as a Markov Decision Process (MDP) and employ the on-policy Proximal Policy Optimization (PPO) algorithm to identify optimal synthetic jet actuation strategies. While PPO and other on-policy methods are widely used in active flow control, off-policy algorithms such as TD3 have also been applied successfully.

The flow control problem is cast as an MDP defined by the tuple $(\mathcal{S}, \mathcal{A}, \mathcal{R}, \gamma)$, where $\mathcal{S}$ is the state space, $\mathcal{A}$ is the action space, $\mathcal{R}$ is the reward function, and $\gamma = 0.99$ is the discount factor.

\vspace{0.2cm}
\noindent 
\textbf{State Space:} The state $s_t \in \mathcal{S}$ at time $t$ consists of pressure measurements extracted from 678 points distributed along the airfoil surface. These pressure values provide a rich representation of the flow field, capturing shock location, boundary layer state, and separation characteristics. The raw pressure values are normalized using
\begin{equation}
s_t^{\text{norm}} = \frac{s_t - \mu(s_t)}{\sigma(s_t) + \epsilon}
\end{equation}
where $\mu(\cdot)$ and $\sigma(\cdot)$ denote the mean and standard deviation, and $\epsilon = 10^{-8}$ prevents division by zero.

\vspace{0.2cm}
\noindent \textbf{Sensor Locations:}
The sensors are randomly positioned on the initial unstructured quadrilateral mesh and remain fixed across all training episodes and experimental configurations to ensure consistency in the state representation and stability in the agent’s observations. Defining sensors on a structured grid would require conservative interpolation to estimate pressure values, which, under AMR, introduces additional computational overhead. To avoid this complexity, the sensor locations are selected in the vicinity of the airfoil from the initial unstructured mesh.

\vspace{0.2cm}
\noindent 
\textbf{Action Space:} The action $a_t \in \mathcal{A}$ represents the velocity scaling factors applied to the synthetic jets distributed along the airfoil surface. Each component $a_t^i$ modulates the jet velocity relative to the freestream velocity $u_\infty$, with bounds $a_t^i \in [-1, 1]$. Positive values correspond to blowing (outflow), while negative values denote suction (inflow) through the jet.

\vspace{0.1cm}
\noindent In particular, we employ the \textit{three-three} jet configuration for effective spatial control. In this setup, the action vector expands to
\[
a_t = [a_t^{\text{top}_1}, a_t^{\text{top}_2}, a_t^{\text{top}_3}, 
       a_t^{\text{bottom}_1}, a_t^{\text{bottom}_2}, a_t^{\text{bottom}_3}]^T,
\]
where each top–bottom jet pair is controlled independently.
All jets are oriented normal to the surface of an airfoil.

\vspace{0.2cm}
\noindent \textbf{Single-agent reinforcement learning}: In both configurations, a single-agent PPO framework is employed. A unified policy network receives the complete set of pressure measurements across the mesh and outputs control signals for all synthetic jets simultaneously. Since shock–boundary layer interaction and aerodynamic efficiency depend on coordinated jet actuation, a single agent is necessary to ensure coherent and coupled decision-making. The top and bottom surface dynamics are intrinsically interdependent; actions on the upper surface influence those on the lower surface and vice versa, thereby modifying the overall pressure distribution and the agent’s state representation. 
In the single-agent PPO formulation, independent mass conservation (for jets) is enforced for both the top and bottom jets, yet no explicit communication between them is required because they share a common global state representation. Conversely, a multi-agent setup would operate under partial observability, where each agent perceives only a subset of the flow field (e.g., top or bottom sensors), thus demanding substantially more training episodes for convergence, which is a major limitation when using computationally intensive CFD solvers.

\vspace{0.2cm}
\noindent 
\textbf{Reward Function:} The reward function $r_t = \mathcal{R}(s_t, a_t)$ is designed to promote drag reduction and lift enhancement. It is defined as
\begin{equation}\label{RewardE}
r_t = w_d \cdot \frac{C_{d,\text{baseline}} - C_{d,t}}{ C_{d,\text{baseline}} + \epsilon} 
+ w_l \cdot \frac{ C_{l,t} - C_{l,\text{baseline}} }{ C_{l,\text{baseline}} +\epsilon} 
- w_Q \left(\frac{Q_t}{\text{max}(Q)}\right)^2 
- w_P \left(\frac{P_t}{\text{max}(P)}\right)^2 
- w_R \frac{|Q_t - Q_{t-1}|}{\text{max}(Q) \Delta t}.
\end{equation}

where $w_d$ and $w_l$ denote the weights associated with drag reduction and lift enhancement, respectively. $C_{d,t}$ and $C_{l,t}$ denotes the instantaneous drag and lift coefficients at time $t$.
The value of $\epsilon$ is typically of order $10^{-4}$--$10^{-6}$. 
The mass flux is represented by $Q$, while the kinetic power of the jet is given by $P  = \tfrac{1}{2} \frac{Q^{3}}{\rho^{2} A_{\mathrm{jet}}^{2}},$
where $\rho$ and $A_{\mathrm{jet}}$ denote the fluid density and the magnitude of jet area, respectively. 
The maximum permissible values of $Q$ and $P$ are predefined according to actuator specifications. 
A quadratic penalty on the control jet amplitude is employed to ensure smooth variations rather than abrupt changes, 
and an additional rate penalty is included to suppress high-frequency actuation (chattering).

\vspace{0.2cm}
\noindent 
\textbf{Control Frequency and Episode Structure:} The RL agent operates in an intermittent control regime with parameters summarized in Table~\ref{tab:rl_params}. The agent reads the flow state and issues control actions every $\Delta t$ second of physical simulation time. During each control interval, the jet velocities remain fixed, creating a piecewise-constant actuation signal. Each episode consists of $N_a$ control steps, corresponding to $T_{\text{ep}}$ seconds of physical flow time. Importantly, the flow field evolves continuously across episode boundaries, the CFD simulation is not restarted, and jet states persist from one episode to the next. This continuous formulation allows the agent to learn control strategies that account for the continuous dynamics of the flow.

\begin{table}
\centering
\caption{Reinforcement learning control parameters}
\label{tab:rl_params}
\begin{tabular}{lc}
\hline
\textbf{Parameter} & \textbf{Value} \\
\hline
Control timestep, $\Delta t$ & 0.01 s \\
Actions per episode, $N_a$ & 50 \\
Episode length, $T_{\text{ep}}$ & 0.5 s \\
Total episodes, $N_{\text{ep}}$ & 200 \\
\hline
\end{tabular}
\end{table}
\begin{figure}[htpb]
    \centering
    \includegraphics[scale=0.63, clip=true]{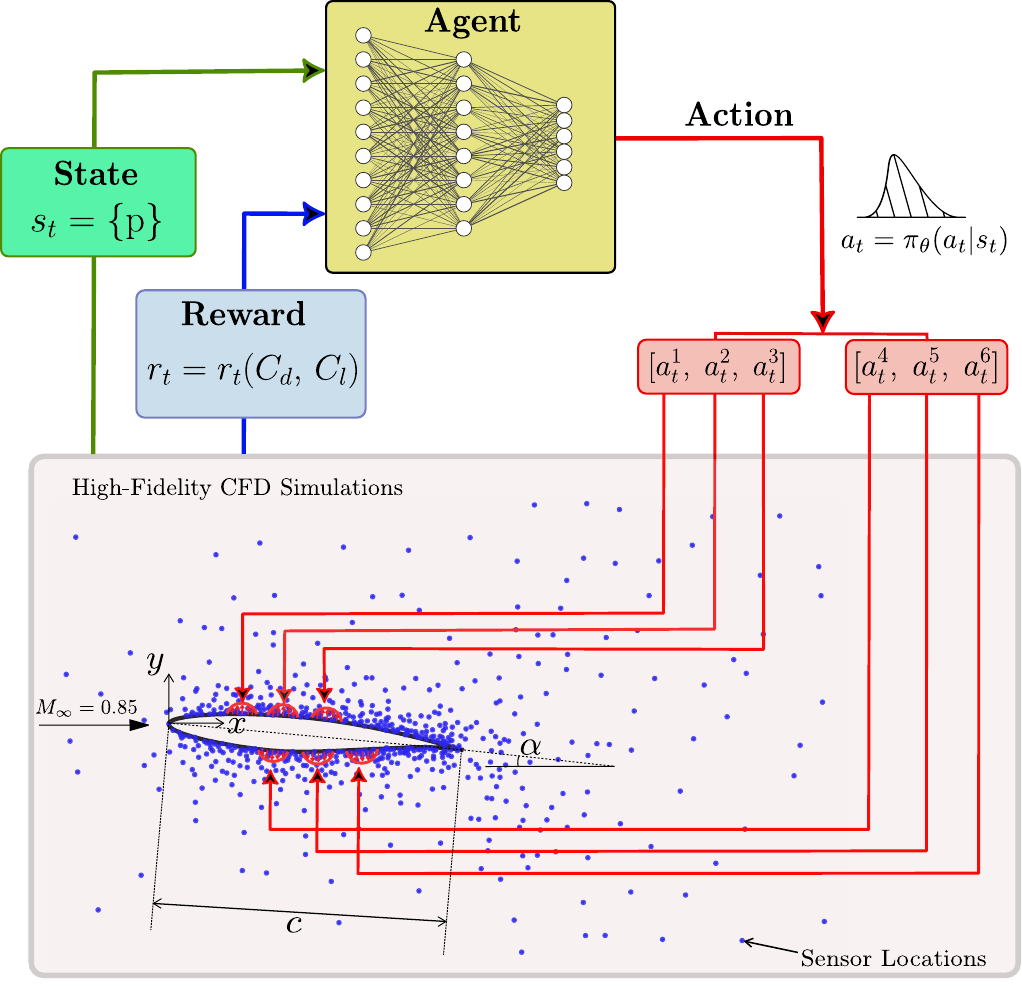}
    \caption{Compressible transonic flow over the RAE2822 2D airfoil with synthetic jet actuation. Three synthetic jets are depicted on both the upper and lower surfaces of the airfoil.}
    \label{fig:RL_airfoil_flowchart} 
\end{figure}
The overall architecture of the DRL--CFD framework employed for transonic flow control over the RAE2822 airfoil using synthetic jet actuators is illustrated in figure \ref{fig:RL_airfoil_flowchart}. The framework couples a high-fidelity CFD solver with a PPO–based learning agent through a closed-loop interaction. At each control step, the CFD solver advances the flow field and provides instantaneous pressure data, which are processed into a state vector, \( \{s_t\} \), and passed to the RL agent. The agent evaluates the current policy, \( \pi_\theta(a_t|s_t) \), to generate control actions corresponding to the synthetic jet mass flow rates. These actions modify the CFD boundary conditions, thereby influencing the subsequent flow evolution. 
The resulting CFD solutions are used to compute a reward, \( r_t \), that reflects improvements in aerodynamic performance by rewarding drag reduction and lift enhancement while penalising excessive jet activity. This reward is supplied to the PPO agent during training, allowing it to progressively discover optimal actuation strategies that balance aerodynamic gains with control effort.

\subsection{Proximal Policy Optimization Algorithm}
\begin{figure}[htpb]
    \centering
    \includegraphics[scale=0.67, clip=true]{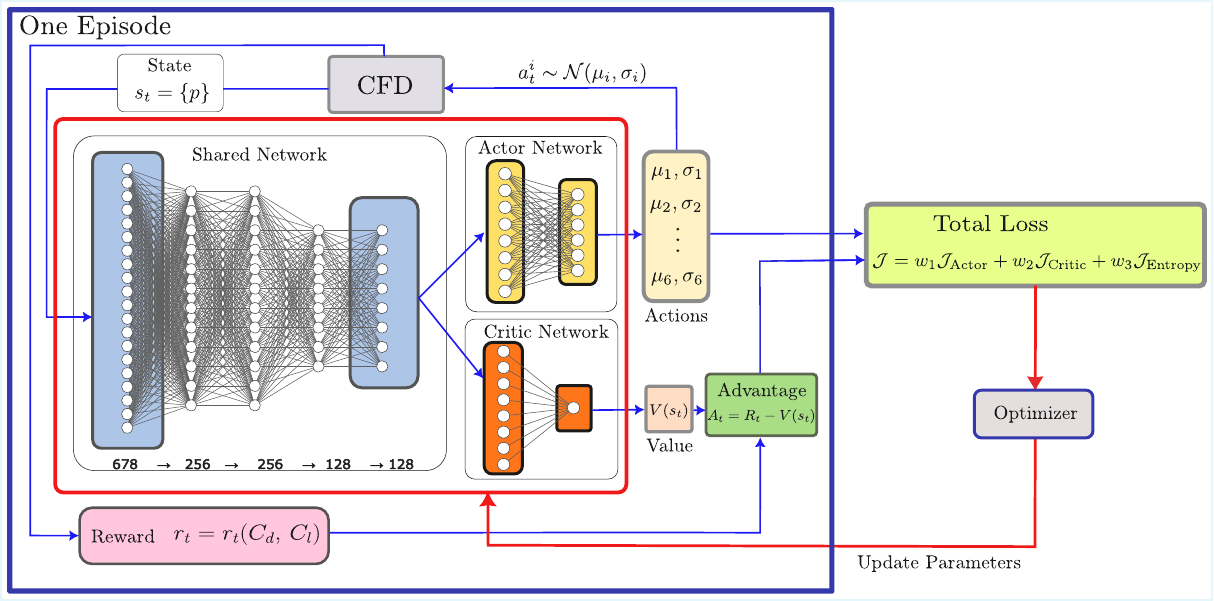}
    \caption{PPO agent's multi-layer perceptron-based neural network architecture used for transonic flow control, where the PPO agent interacts with CFD solver to provide optimal controls for synthetic jet actuations.}
    \label{fig:PPO} 
\end{figure}

The PPO algorithm is employed to train the reinforcement learning agent that determines the jet control strategy from the instantaneous aerodynamic state. PPO follows an \textit{actor–critic} framework, where both the actor (policy) and critic (value estimator) share a common network to efficiently encode the flow state. Figure~\ref{fig:PPO} shows the PPO architecture used in this study. The input layer receives the normalized pressure distribution from 678 sensor points on the airfoil surface, forming a 678-dimensional state vector. This vector is processed through four shared hidden layers, followed by distinct single-layer heads for the actor and critic. The shared layers have dimensions \(256 \rightarrow 256 \rightarrow 128 \rightarrow 128\) (representing the four hidden-layers in the feed-forward neural network). An adaptive activation function, denoted \texttt{Adaptive ReLU} (\cite{jagtap2020adaptive}), is employed and defined as \(\text{ReLU}(10 \cdot a \cdot x)\), where \(a\) is a single global learnable parameter initialized to 0.1. The network is optimized using the Adam optimizer with a learning rate of \(10^{-4}\). The resulting latent representation is split into two heads: the \textit{actor}, which maps state to corresponding actions for the synthetic jet control, and the \textit{critic}, which estimates the state value from a given state.
In the DRL–CFD loop, the PPO network receives flow states from the CFD solver at each control step, generates jet control actions, and the solver advances the flow for the next time interval ($\Delta t$). After each episode (50 control steps), the PPO agent updates its policy based on the collected state–action–reward data, closing the learning loop between the CFD solver and RL controller.

The \textit{actor head} maps the 128-dimensional latent vector to a six-dimensional mean action $\mu_\theta(s_t)$ and the corresponding six-dimensional standard deviation vector ($\sigma$), which is a separate learnable parameter vector that is optimized during training along with networks parameters, but does not depend on the input state. Then the $\tanh$ activation function is used to confine outputs to $[-1, 1]$, matching the physical jet actuation limits. The stochastic policy is modeled as a Gaussian distribution,
\[
\pi_\theta(a_t|s_t) = \mathcal{N}(\mu_\theta(s_t), \sigma),
\]
where $\sigma$ is a learnable standard deviation governing exploration.

The \textit{critic head} predicts a scalar value $V(s_t)$, representing the expected cumulative reward from state $s_t$, and serves as a baseline to reduce policy-gradient variance. The discounted return is
\[
R_t = \sum_{k=0}^{\infty}  \gamma^k r_{t+k} 
\]
and the advantage, $A_t = R_t - V(s_t),$
measures how much better or worse the action performed than expected. Positive $A_t$ values increase the likelihood of similar actions, while negative values suppress them.

\vspace{0.2cm}
\noindent \textbf{Actor Loss (Clipped Surrogate Objective) :}
The actor network is optimized using the clipped surrogate objective:
\[
\mathcal{J}_{\text{actor}} = -\mathbb{E}_t \left[
\min\left(
r_t(\theta) A_t,\;
\text{clip}(r_t(\theta), 1 - \epsilon, 1 + \epsilon) A_t
\right)
\right],
\]
where
\[
r_t(\theta) = \frac{\pi_\theta(a_t|s_t)}{\pi_{\text{old}}(a_t|s_t)}
\]
is the probability ratio between the new and old policies, and $\epsilon$ is the clipping parameter (in our case, it’s $0.2$). The negative sign ensures that gradient descent minimizes the negative of the clipped objective, effectively maximizing the PPO surrogate function. This clipping mechanism limits large updates to the policy and stabilizes training by preventing the policy from deviating too far from its previous version.

\vspace{0.2cm}
\noindent \textbf{Critic Loss :}
The critic is trained by minimizing the mean squared error between its value prediction and the discounted return:
\[
\mathcal{J}_{\text{critic}} = \mathbb{E}_t\big[(R_t - V(s_t))^2\big].
\]
This term ensures that the critic accurately estimates the expected future rewards, which provides a baseline for computing advantages.

\vspace{0.2cm}
\noindent \textbf{Entropy Regularization :}
To encourage exploration and prevent the policy from collapsing into a deterministic behavior too early in training, an entropy regularization term is introduced. The entropy loss for the policy distribution is given by:
\[
\mathcal{J}_{\text{entropy}} = -\mathbb{E}_t[\log \pi_\theta(a_t|s_t)].
\]
Higher entropy corresponds to greater stochasticity in action selection. In implementation, this loss is with a negative coefficient, so that maximizing the overall PPO objective indirectly maximizes entropy and encourages exploration.

\vspace{0.2cm}
\noindent \textbf{Total PPO Loss :}
The overall PPO loss combines the three components: actor, critic, and entropy regularization into a single objective:
\[
\mathcal{J}_{\text{total}} =
w_1 \mathcal{J}_{\text{actor}}
+ w_2 \mathcal{J}_{\text{critic}}
+ w_3 \mathcal{J}_{\text{entropy}},
\]
where $w_1 = 1.0$, $w_2 = 0.5$, and $w_3 = 0.01$,  are weights that balance the contributions of policy optimization, advantage estimation, and exploration, respectively.  
Both the actor and critic networks are trained jointly using gradient descent to minimize this total loss.

\section{Results}
We examine the three–three jet arrangement, which consists of pairs of jets distributed along the upper and lower surfaces of the airfoil. This configuration is designed to ensure that the number and placement of jets are sufficient to exert a meaningful influence on the surrounding flow field. By appropriately energizing the boundary layer and modulating the shock–boundary-layer interaction, the jets can enhance aerodynamic performance.

Unless otherwise specified, the DRL-based controller is trained using the reward function given by \eqref{RewardE}, with $w_d = 2.0$ and $w_l = 1.0$, reflecting the prioritization of drag reduction over lift enhancement. This choice is particularly relevant for cruise flight conditions (e.g., passenger aircraft operating near transonic Mach numbers), where minimizing drag directly improves fuel efficiency and overall aerodynamic performance.

\subsection{Three-Three Jets}
This configuration aims to provide greater controllability and improved influence over the flow field. The jets are positioned at the following chord-wise locations: on the top surface at $0.2c$–$0.3c$, $0.35c$–$0.45c$, and $0.5c$–$0.6c$, and on the bottom surface at $0.35c$–$0.45c$, $0.5c$–$0.6c$, and $0.65c$–$0.75c$.

\begin{figure}[H]
    \centering
    \includegraphics[width=1.0\textwidth]{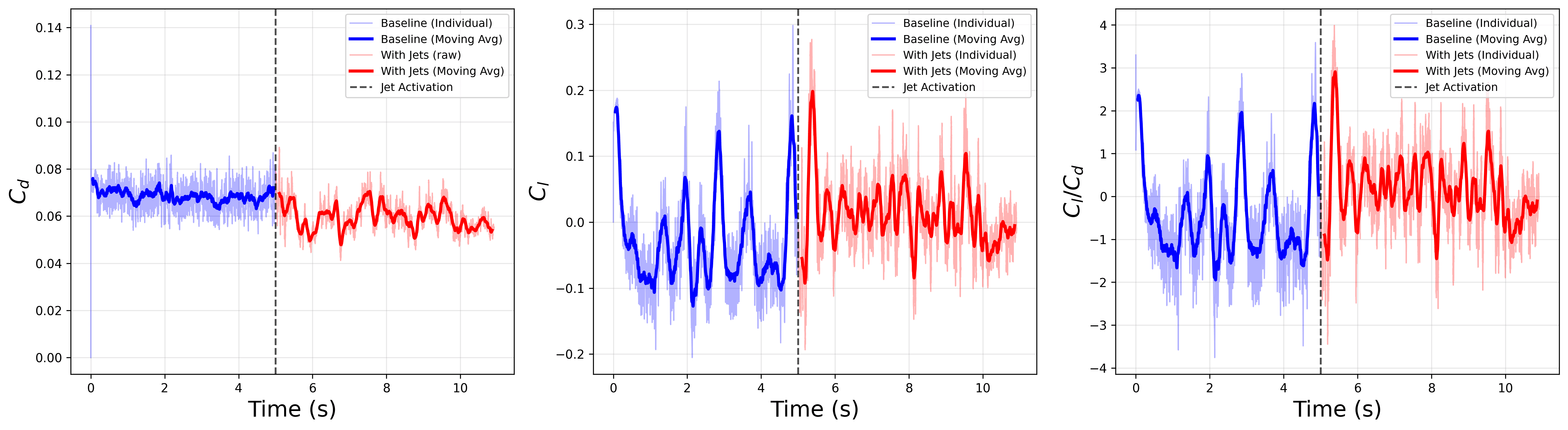}
    \caption{Temporal evolution of aerodynamic coefficients:  From left to right, the panels show the drag coefficient, lift coefficient and aerodynamic efficiency as functions of time. The blue line denotes the baseline flow, while the red line corresponds to the flow with jet actuation.}
    \label{fig:three_jets_cl_cd}
\end{figure}

\begin{figure}[H]
    \centering
    \includegraphics[scale=0.42, clip=true]{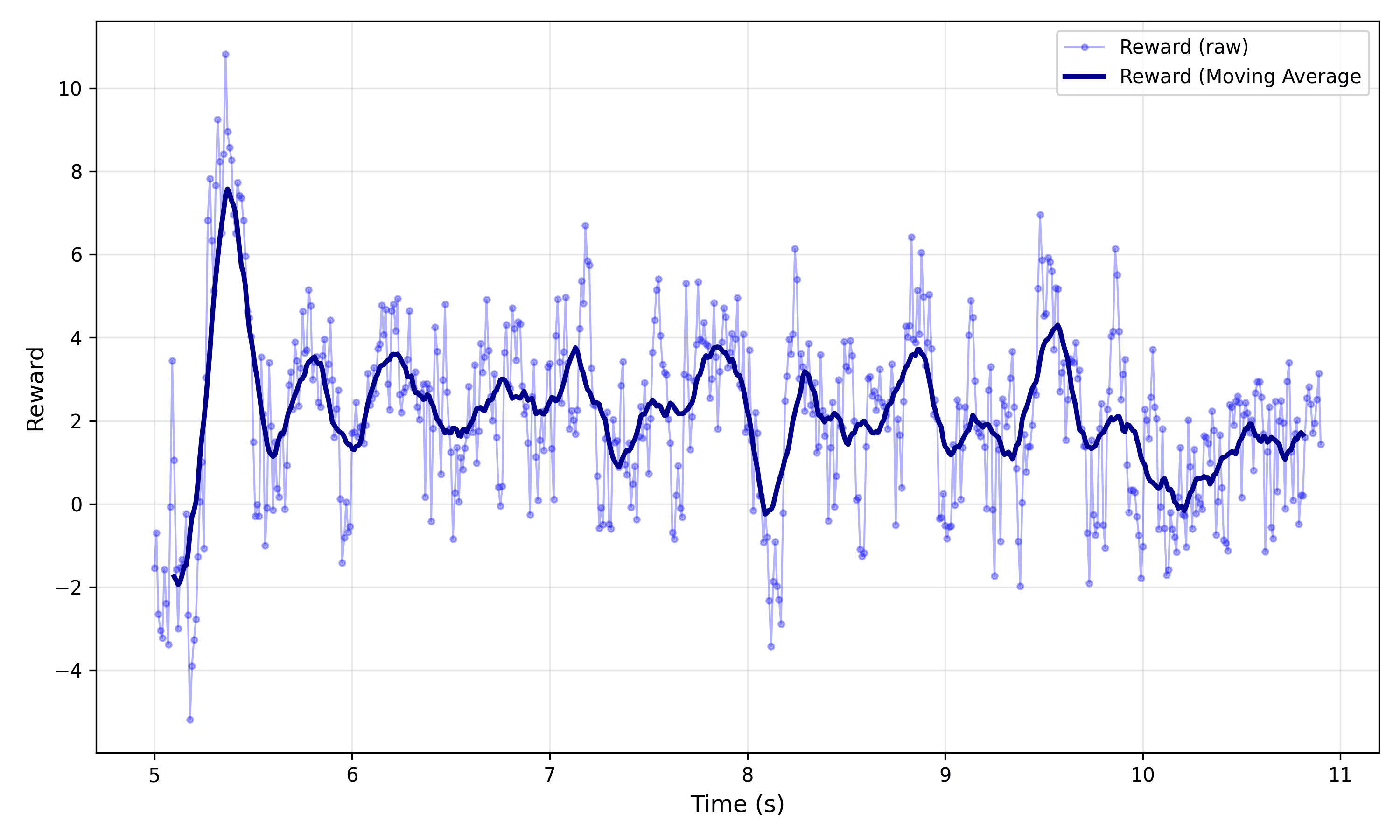}
    \caption{Training performance of the three–three jet configuration: Shown is the temporal evolution of the reward function during training.}
    \label{fig:three_jets_reward_time}
\end{figure}

\begin{figure}[htpb]
    \centering
    \includegraphics[scale=0.07, clip=true]{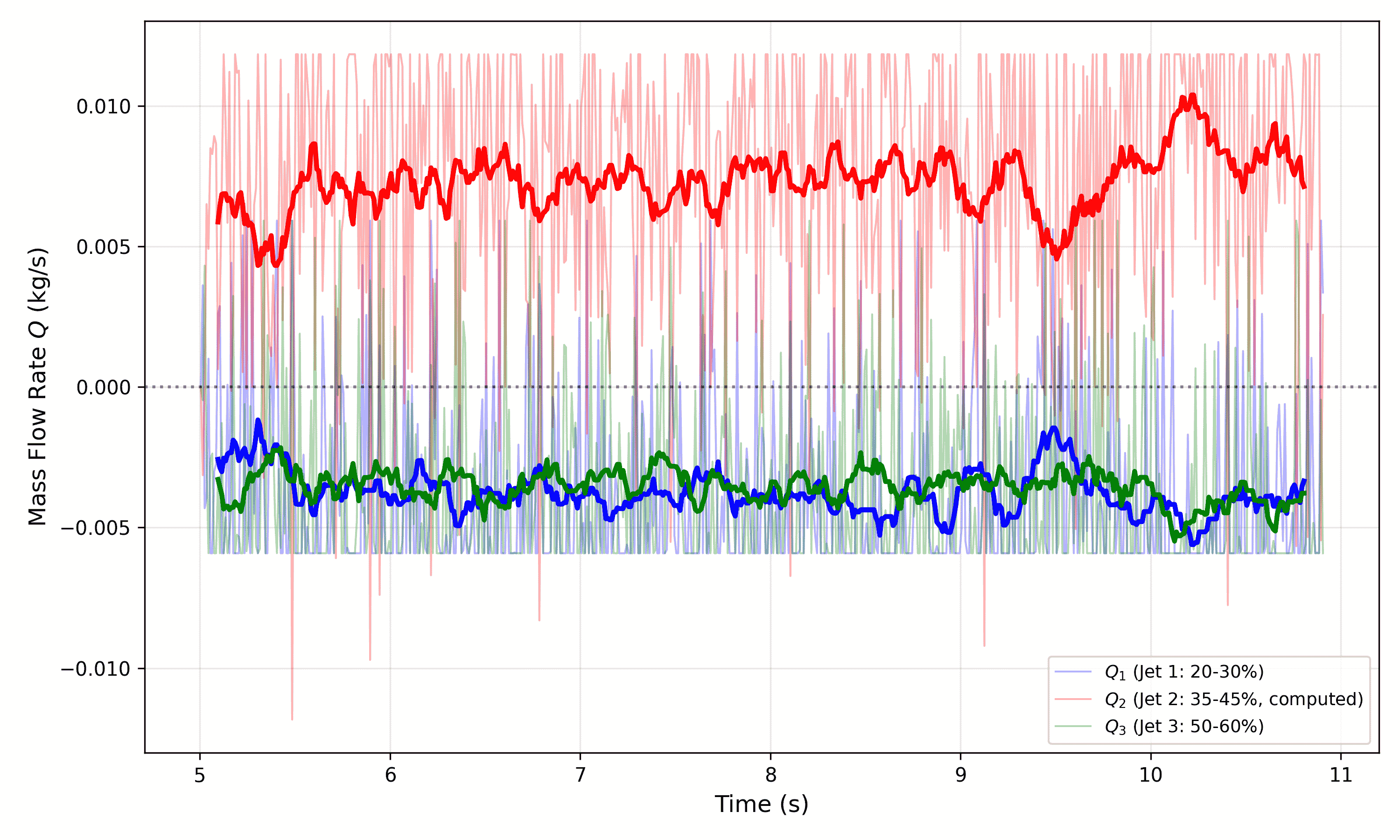}
        \includegraphics[scale=0.07, clip=true]{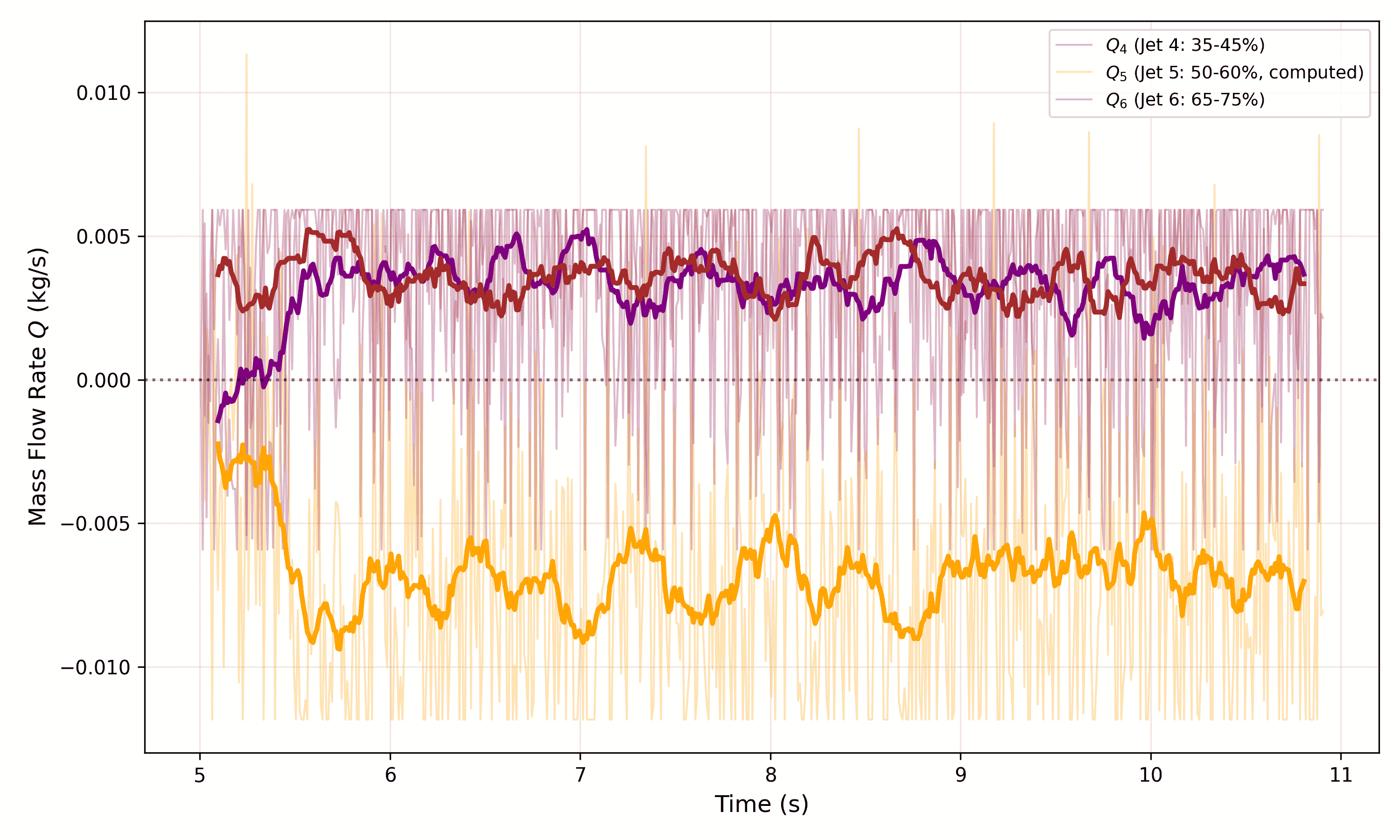}
    \caption{Mass flow rate evolution for top-surface jets (Q1-Q3) and bottom-surface jets (Q4-Q6) in the three–three jet configuration.}
    \label{fig:three_jets_mass_flow_rate}
\end{figure}

Again, to maintain overall mass conservation across the airfoil, the central jet on each surface is defined as a function of the two outer jets: $Q_2 = -\left(Q_1 + Q_3\right), $ and $ Q_5 = -\left(Q_4 + Q_6\right).$
This results in a four-dimensional action space.
 The RL agent for this configuration uses the same reward function formulation as in the two-two case. For the training setup, the control time step is $\Delta t = 0.01 \text{s}$, and each episode spans a duration of $T_{\text{ep}} = 0.5~\text{s}$, corresponding to $N_a = 50$ discrete control actions per episode. The combined aerodynamic coefficient evolution for this case is shown in Figure \ref{fig:three_jets_cl_cd} and the variation of the reward function with time during training is presented in Figure \ref{fig:three_jets_reward_time}. The mass flow rate and its mean values (moving average) evolution for individual jets on the top and bottom surfaces is shown in Figure \ref{fig:three_jets_mass_flow_rate}. Here, Q1-Q3 correspond to the upper-surface jets and Q4-Q6 to the lower-surface jets.

A comparison between the baseline and actuated flow conditions is summarized in Table \ref{tab:three_three_results}. The results shows a clear improvement in aerodynamic efficiency compared to the two-two jet configuration. Upon actuation, the average lift coefficient ($C_l$) increases significantly by $131.18\%$, while the drag coefficient ($C_d$) decreases by $13.78\%$.
 The combined lift-to-drag ratio improves by approximately $121.52\%$, which shows the beneficial influence of this more distributed jet arrangement. 
\begin{table}
\centering
\caption{Comparison of average aerodynamic coefficients for the three-three jet configuration}
\label{tab:three_three_results}
\begin{tabular}{lcc}
\hline
\textbf{Parameter} & \textbf{Baseline} & \textbf{With Jets} \\
\hline
Average $C_d$ & 0.068768 & 0.066055 \\
Average $C_l$ & -0.028659 & 0.175099 \\
Change in $C_d$ & \multicolumn{2}{c}{$-13.78\%$} \\
Change in $C_l$ & \multicolumn{2}{c}{$+131.18\%$} \\
\hline
\end{tabular}
\end{table}
These observations indicate that the set of three jets on each surface enhances the controller’s ability to manipulate the flow field more effectively. The resulting flow actuation not only improves lift but also contributes to drag reduction, which means better controllability.

\subsection{Three-Three Jets with Distributed Environments}

\begin{figure}[htpb]
    \centering \includegraphics[scale=0.65, clip=true]{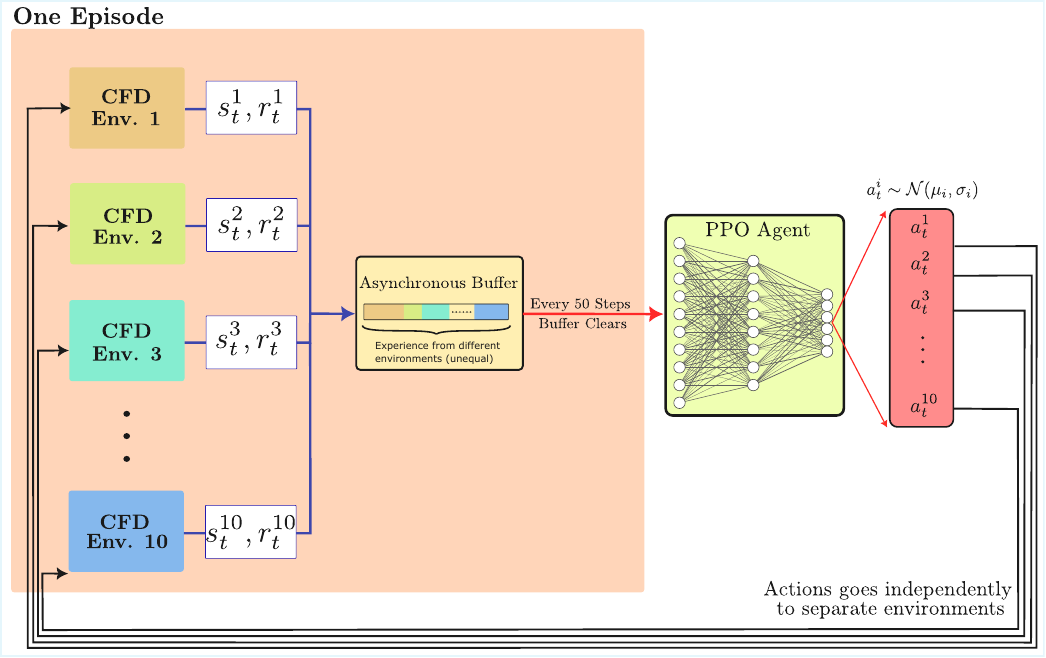}
    \caption{Schematic of the distributed framework where independent environments collect experiences and store them in a shared asynchronous buffer.}
    \label{fig:parallel_env}
\end{figure}

To further accelerate training convergence, exploration, and improve policy generalization, the three-three jet configuration is extended to a distributed training framework using multiple asynchronous environments (Figure \ref{fig:parallel_env}). All CFD simulations and PPO training were executed entirely on CPU resources. Each CFD environment was run on 16 CPU cores, with a total of 170 cores utilized concurrently across distributed environments. The overall training process spanned approximately 40 wall-clock hours, corresponding to nearly 6,800 CPU-core hours of computation. Both the CFD solver and the RL framework were efficiently coupled, with each CFD environment, policy inference, and policy update all running in parallel.

Each environment simulates the same airfoil configuration with different actions sampling from a common PPO agent. This multi-environment setup ensures broader policy generalization and prevents overfitting to a single flow environment. 
In this distributed framework, each environment runs its own simulation and collects experiences independently. These experiences are pushed into a shared asynchronous buffer, which acts as a global memory for all environments. Once the buffer reaches a predefined capacity (in this case, every 50 steps), the policy and value networks are updated using the aggregated experiences, and the buffer is cleared. This policy updates through asynchronous data collection allows to have more training experiences, which enhances policy stability.

\begin{figure}[htpb]
    \centering
    \includegraphics[width=1.0\textwidth]{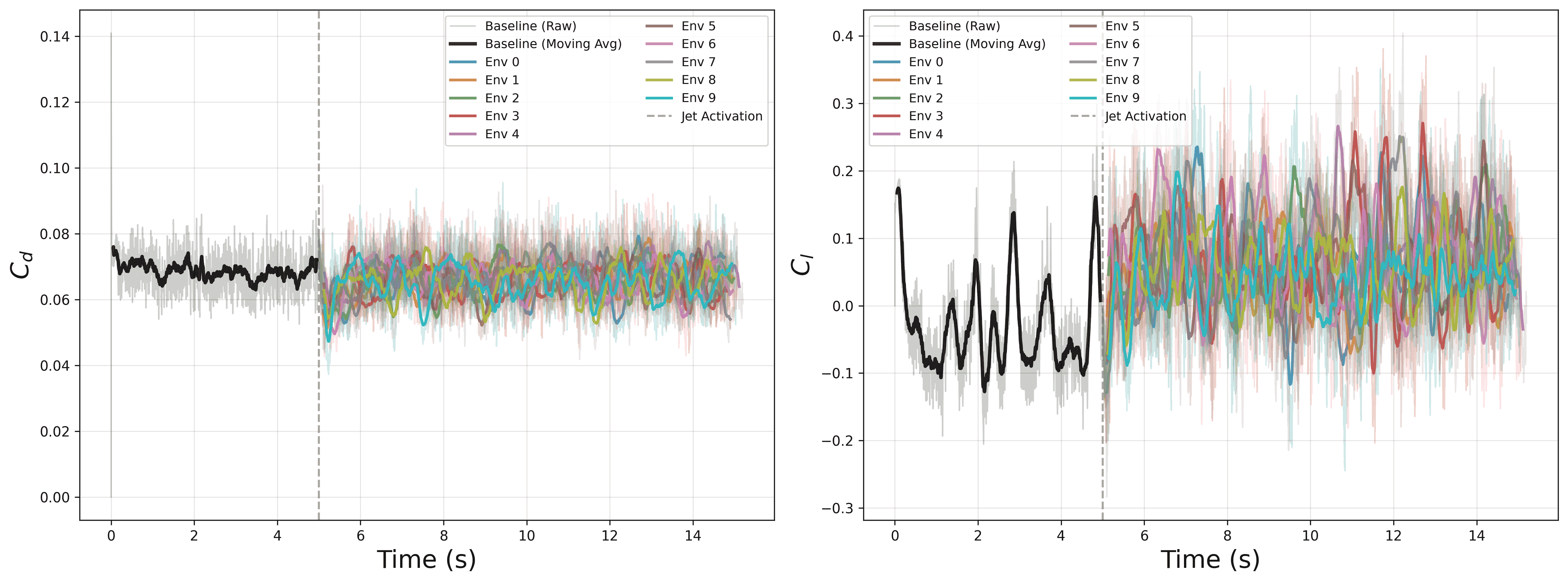}
    \caption{Left: drag coefficient as a function of time. Right: lift coefficient as a function of time. The black line denotes the baseline flow, the vertical dashed line indicates the onset of jet activation, and the different environments are shown in different colors.}
    \label{fig:multi_three_jets_cl_cd}
\end{figure}

\begin{figure}[htpb]
    \centering
    \includegraphics[scale=0.35, clip=true]{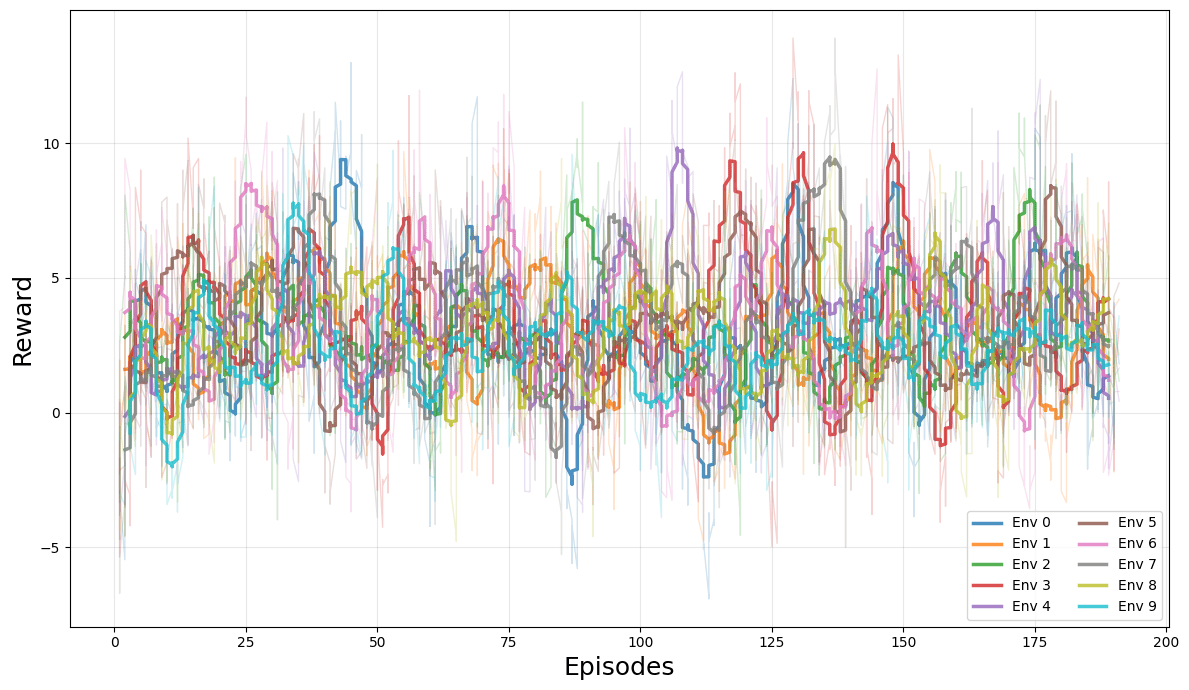}
    \caption{Reward vs episodes for the three–three jet configuration using different environments (Env's).}
    \label{fig:multi_three_jets_reward_episode}
\end{figure}

\begin{figure}[htpb]
    \centering
    \includegraphics[scale=0.20, clip=true]{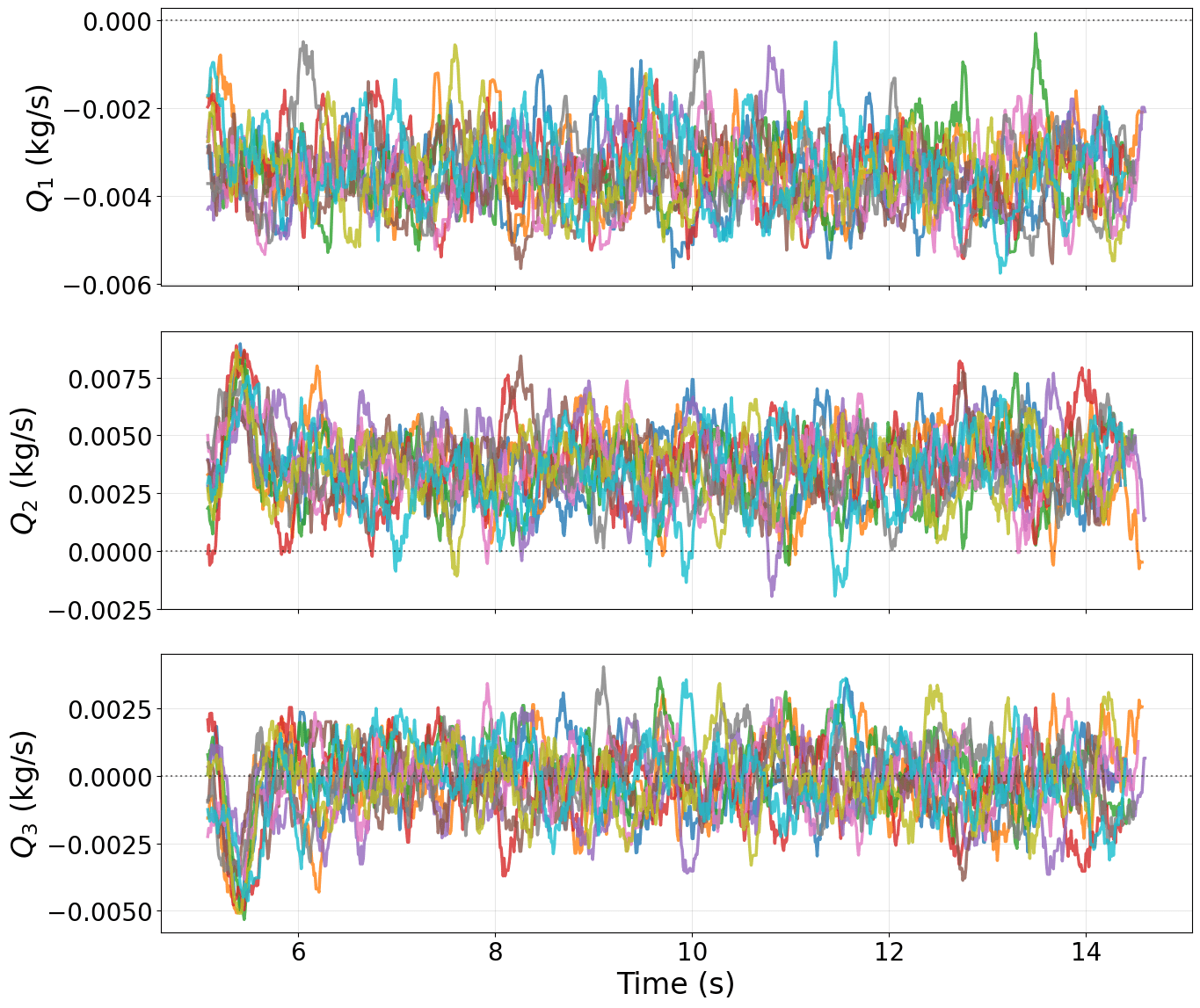}
    \includegraphics[scale=0.20, clip=true]{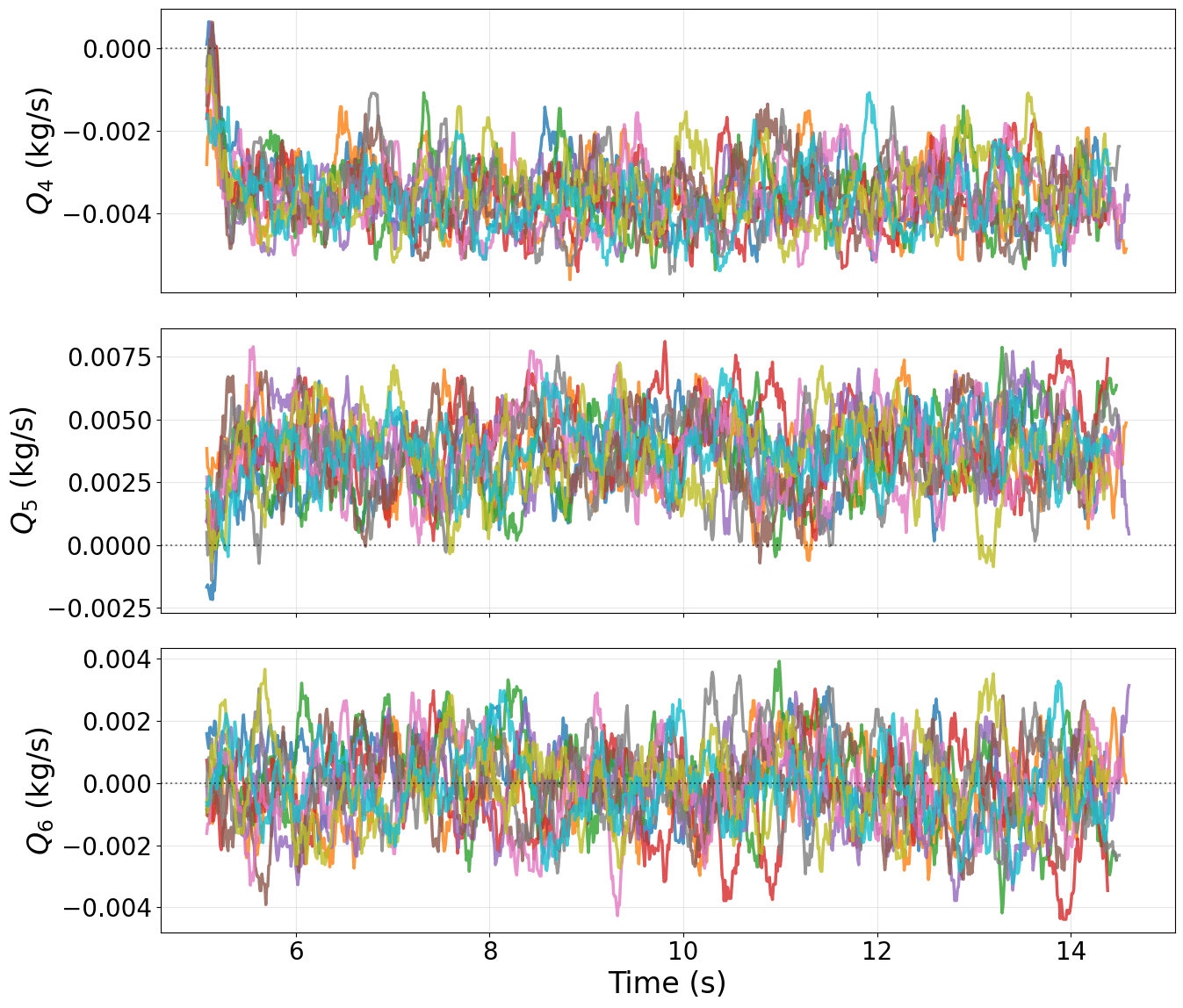}
    \includegraphics[scale=0.23, clip=true]{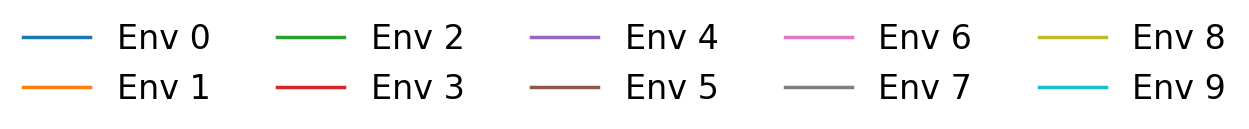}
    \caption{Mass flow rate evolution for top-surface jets (Q1-Q3) and bottom-surface jets (Q4-Q6) in the three–three jet configuration.}
    \label{fig:multi_three_jets_mass_flow_rate}
\end{figure}

Following jet activation at $t \approx 5.0~\mathrm{s}$, a distinct reduction in $C_d$ is observed across all environments (Figure~\ref{fig:multi_three_jets_cl_cd}). Here we report the best result among all environments. The drag subsequently stabilises around a lower mean value with moderate oscillations, corresponding to an overall reduction of approximately $7.50,\%$ relative to the baseline. Concurrently, the $C_l$ increases markedly by about $160.19,\%$, accompanied by oscillations due to flow dynamics, thereby enhancing the overall aerodynamic performance. The evolution of the reward function during training (Figure~\ref{fig:multi_three_jets_reward_episode}) shows a steady increase, with the cumulative reward rising from less than $-2.5$ to approximately $2.5$ as learning progresses. This trend reflects the gradual convergence of the PPO agent towards an effective and stable jet control strategy. The mass flow rate distributions (Figure~\ref{fig:multi_three_jets_mass_flow_rate}) exhibit consistent phase-synchronised behaviour among the upper and lower jets, indicating that the PPO agents have learned coordinated actuation across distributed control units. On the upper surface, the front and middle jets predominantly operate in continuous blowing and suction modes, respectively, while similar behaviour is observed on the lower surface, with the front jet mainly blowing and the middle jet maintaining suction.

\subsubsection{Different Actuation Frequency}
To examine the influence of actuation frequency on aerodynamic performance, another study is conducted using four distinct actuation frequencies. The objective is to assess how variations in jet modulation rate affect the lift and drag characteristics of the airfoil under the same three–three jet configuration. Here, $f_s$ denotes the dominant natural frequency of the system (unactuated flow), determined from spectral analysis of the unforced dynamics and found to be approximately $34~\mathrm{Hz}$. The actuation frequency, $f_a$, corresponds to the rate at which control actions are applied by the RL agent, representing the temporal update of the synthetic jets. Four actuation frequency ratios, $f_a/f_s = 1, 3, 10, \text{and},50$, are investigated, corresponding to control time intervals of $\Delta t = 0.03$, $0.01$, $0.003$, and $0.0006~\mathrm{s}$, respectively, while maintaining a constant number of actions per episode ($N_a = 50$). This parameter range permits the analysis of both low- and high-frequency actuation regimes. It is important to note that all three–three jet results discussed previously were obtained at $f_a/f_s = 3$.

\begin{table}
\centering
\caption{Effect of actuation frequency ratio $f_a/f_s$ on aerodynamic performance.}
\begin{tabular}{c c c}
\hline
\textbf{$f_a/f_s$} & \textbf{$\Delta C_d^{\text{best}}$} & \textbf{$\Delta C_l^{\text{best}}$} \\
\hline
1  & -6.23\%  & +188.20\% \\
3  & -7.50\% & +160.19\% \\
10 & -3.87\% & +109.95\% \\
50 & -3.20\% & -168.02\% \\
\hline
\end{tabular}
\label{tab:actuation_frequency}
\end{table}

The comparative performance across these cases is summarized in Table~\ref{tab:actuation_frequency}. The moderate actuation frequency ratios ($f_a / f_s = 1$ and $3$) yield the best aerodynamic performance, achieving approximately $188.20\%$ and $160.19\%$ increases in lift and $6.23\%$ and $7.50\%$ reductions in drag relative to the baseline, respectively.
 These results indicate the existence of an optimal actuation frequency that enables the controller to synchronize most effectively with the unsteady flow dynamics and maximize aerodynamic efficiency. 
 Conversely, at very high frequency ratios ($f_a / f_s = 10$ and above), the rapid actuation diminishes flow sensitivity and controllability, as the flow cannot respond effectively to such fast forcing. The lift coefficients remain negative throughout the simulation, particularly for $f_a / f_s = 50$, indicating that excessive forcing disrupts rather than enhances aerodynamic performance. This suggests that beyond a certain frequency threshold, the actuation energy is dissipated without contributing to effective flow control, thereby degrading overall performance.

\subsection{Reward Function for Drag Reduction and Lift Preservation}
Using the same distributed training environments, a new set of experiments is conducted by reformulating the reward function. While previous studies focused primarily on drag reduction and lift enhancement, the present formulation aims to minimize drag while preserving a stable lift coefficient, as typically required under cruise conditions for commercial passenger aircraft. This remains a significant challenge for DRL-based control, as the flow dynamics are highly unsteady and strongly influenced by shock–boundary layer interactions. The revised reward function retains the structure of Equation~\ref{RewardE}, except for the lift-related term and its associated weighting factors:
\begin{equation}\label{RewardE_2}
r_t = w_d \cdot \frac{C_{d,\text{baseline}} - C_{d,t}}{C_{d,\text{baseline}} + \epsilon} 
- w_l \cdot \frac{\left| C_{l,t} - C_{l,\text{baseline}} \right|}{\left| C_{l,\text{baseline}} + \epsilon \right|} 
- w_Q \left( \frac{Q_t}{\text{max}(Q)} \right)^2 
- w_P \left( \frac{P_t}{\text{max}(P)} \right)^2 
- w_R \frac{|Q_t - Q_{t-1}|}{\text{max}(Q)\, \Delta t}.
\end{equation}

Two sets of weight combinations are evaluated, as summarized in Table~\ref{tab:reward_weights}. The results are shown for $w_d = 2.0, w_l = 0.1$ and $w_d = 2.0, w_l = 1.0$, which perform similarly. This outcome highlights the inherent difficulty of reducing drag while preserving lift in transonic flow, where the two quantities are strongly coupled through shock–boundary-layer interactions. Drag reduction often requires modifying shock strength or delaying boundary-layer separation, which directly affects the pressure distribution and, consequently, lift. Unlike subsonic regimes, small actuation changes in transonic flows can induce large, nonlinear lift fluctuations through shock–boundary layer interactions. The agent’s performance therefore reflects this intrinsic aerodynamic coupling rather than a limitation of the reward formulation. This highlights that effective DRL control in transonic regimes must exploit subtle flow manipulations that simultaneously respect drag–lift coupling.

\begin{table}
\centering
\caption{Effect of reward function modification on drag reduction and lift preservation.}
\label{tab:reward_weights}
\begin{tabular}{ccc}
\hline
$(w_d,\, w_l)$ & \textbf{$\Delta C_d^{\text{best}}$} & \textbf{$\Delta C_l^{\text{best}}$} \\
\hline
(2.0, 1.0) & $-4.45\%$ & $+45.08\%$ \\
(2.0, 0.1) & $-3.60\%$ & $+19.52\%$ \\
\hline
\end{tabular}
\end{table}

\subsection{Comparison Between Off-Policy and On-Policy DRL}
Next, we compare active flow control performance using two DRL approaches: the on-policy PPO and the off-policy Twin Delayed Deep Deterministic Policy Gradient (TD3) by \cite{fujimoto2018addressing}. This comparison highlights how distinct learning paradigms influence control stability, convergence behaviour, and sample efficiency. On-policy methods, such as PPO, learn exclusively from data generated by the current policy, requiring fresh environment interactions after each update. This approach results in low bias but high variance, as past experiences cannot be reused. Despite lower sample efficiency, on-policy algorithms often exhibit robust and stable training in complex continuous-control tasks due to the consistency of their data distribution.
By contrast, off-policy methods, such as Deep Deterministic Policy Gradient (DDPG) by \cite{lillicrap2015continuous} and TD3, learn from experiences generated by any policy stored in a replay buffer. This enables extensive reuse of past samples, markedly improving sample efficiency. However, because the replayed data may not perfectly reflect the current policy’s state–action distribution, off-policy learning introduces bias into value estimates and gradient updates, which can compromise stability and convergence.
\begin{table}
\centering
\caption{Results: PPO vs. TD3}
\label{tab:td3_vs_ppo}
\begin{tabular}{ccc}
\hline
\textbf{Methods} & \textbf{$\Delta C_d^{\text{best}}$} & \textbf{$\Delta C_l^{\text{best}}$} \\
\hline
On-Policy (PPO) & $-4.45\%$ & $+45.08\%$ \\
\textbf{Off-Policy (TD3)} & $\textbf{-25.62\%}$ & $\textbf{+196.30\%}$ \\
\hline
\end{tabular}
\end{table}
\begin{figure}[htpb]
    \centering
    \includegraphics[width=0.95\textwidth]{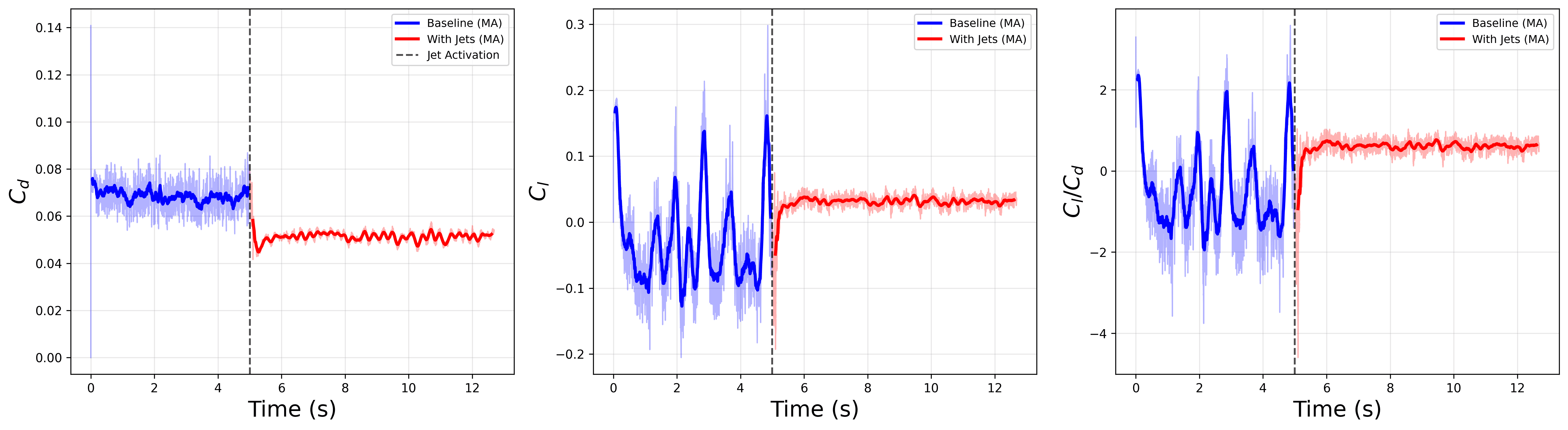}
    \caption{Temporal evolution of aerodynamic coefficients for TD3:  From left to right, the panels show the drag coefficient, lift coefficient and aerodynamic efficiency as functions of time. The blue line denotes the baseline flow, while the red line corresponds to the flow with jet actuation. MA stands for 'moving average'.}
    \label{fig:td3_cl_cd}
\end{figure}
\begin{figure}[htpb]
    \centering
    \includegraphics[scale=0.4, clip=true]{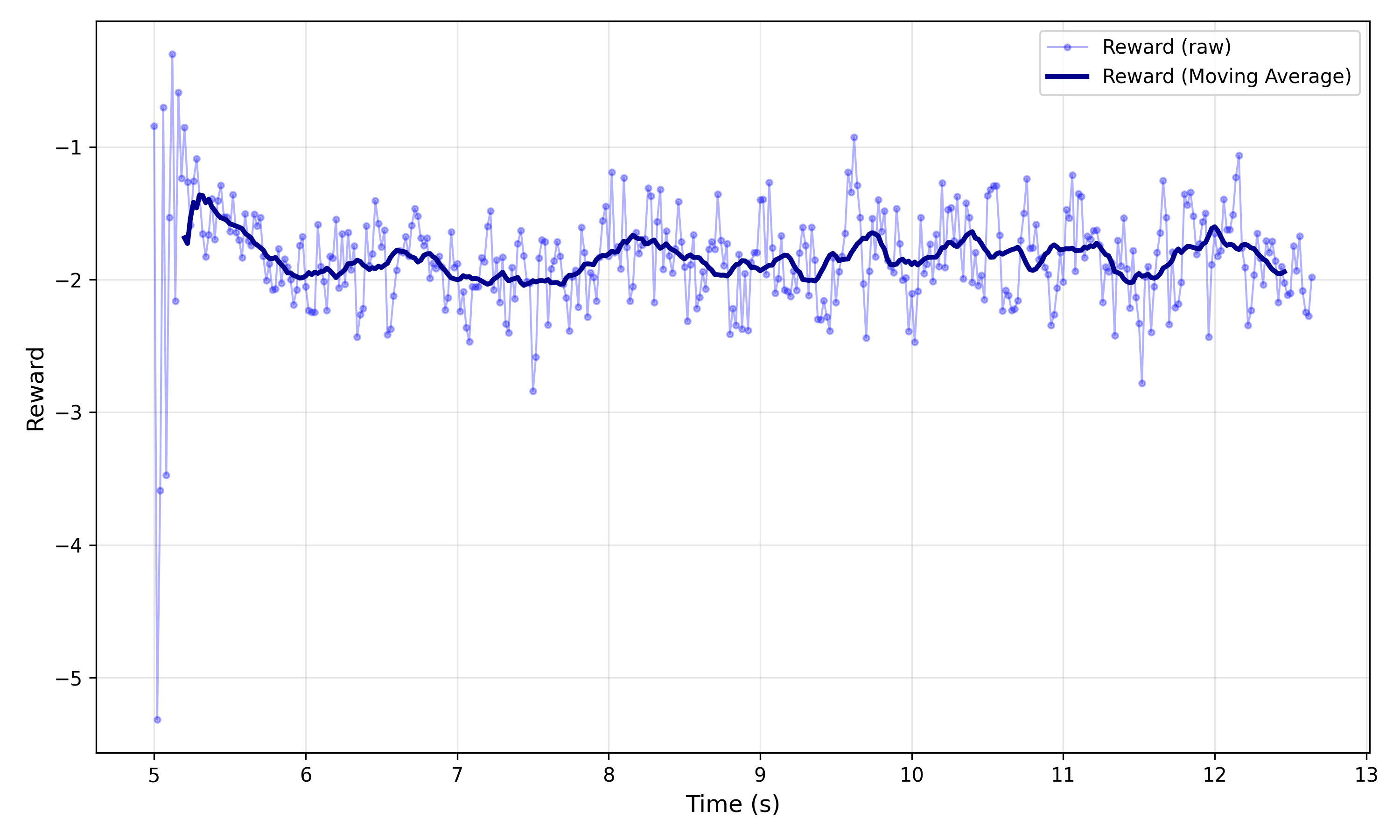}
    \caption{Temporal evolution of the reward function: Shown is the variation of the reward during off-policy TD3 training.}
    \label{fig:td3_reward}
\end{figure}
\begin{figure}[htpb]
    \centering
    \includegraphics[scale=0.21, clip=true]{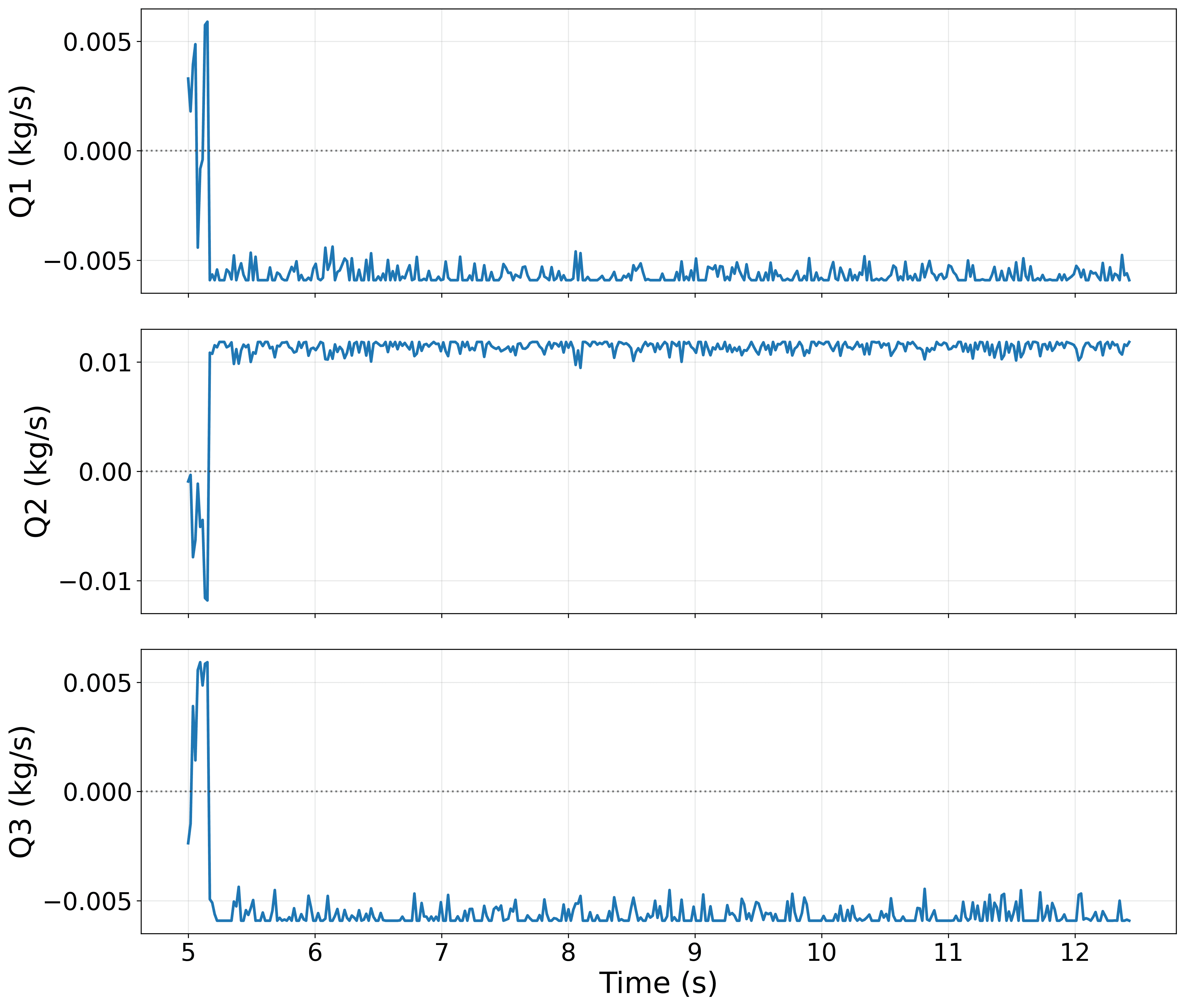}
    \includegraphics[scale=0.21, clip=true]{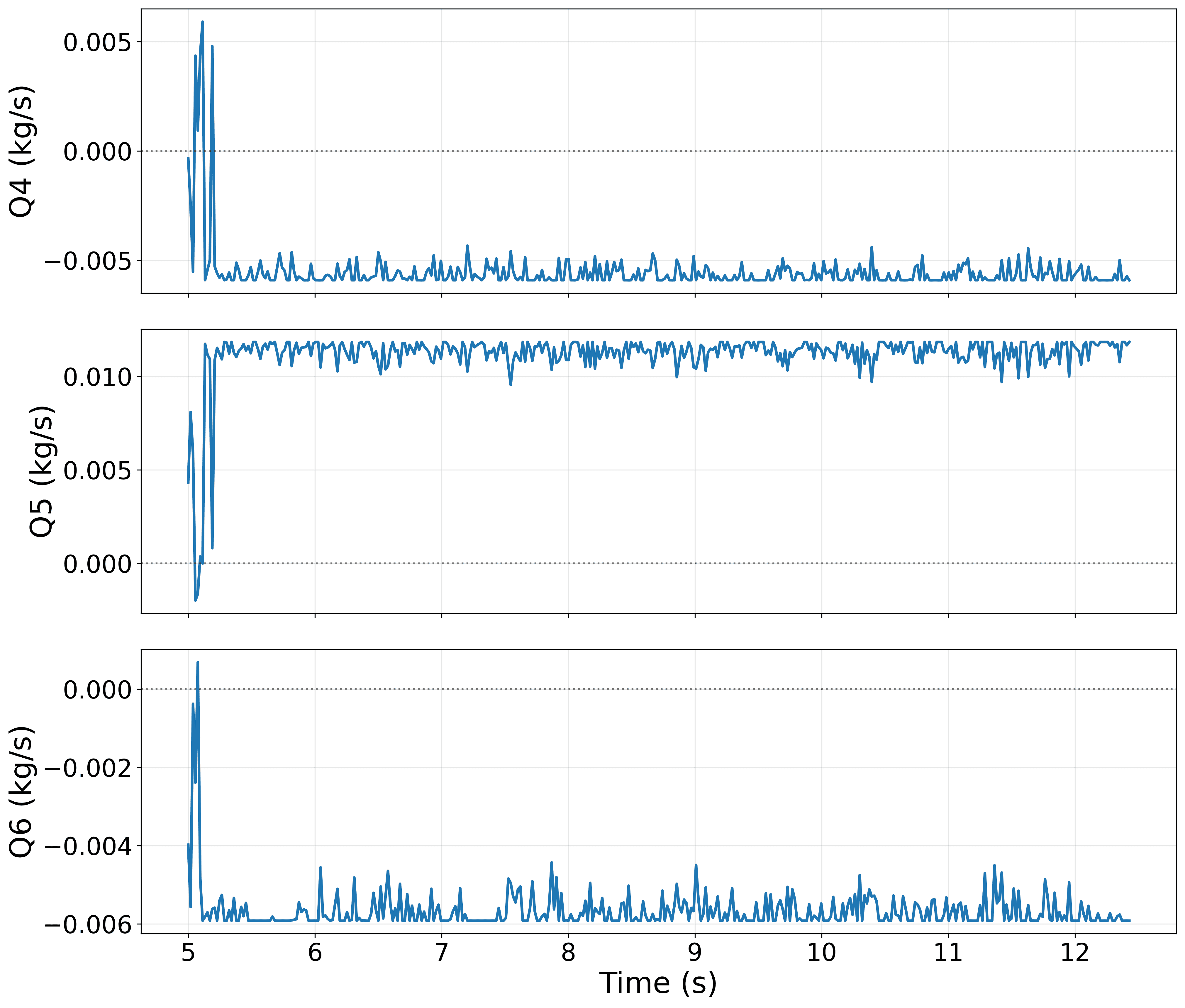}
    \caption{Evolution of mass flow rate for top-surface (Q1-Q3) jets  and bottom-surface (Q4-Q6) jets during TD3 training.}
    \label{fig:td3_mass_flow_rate}
\end{figure}
\begin{figure}[htpb]
    \centering
    \includegraphics[scale=0.83, clip=true]{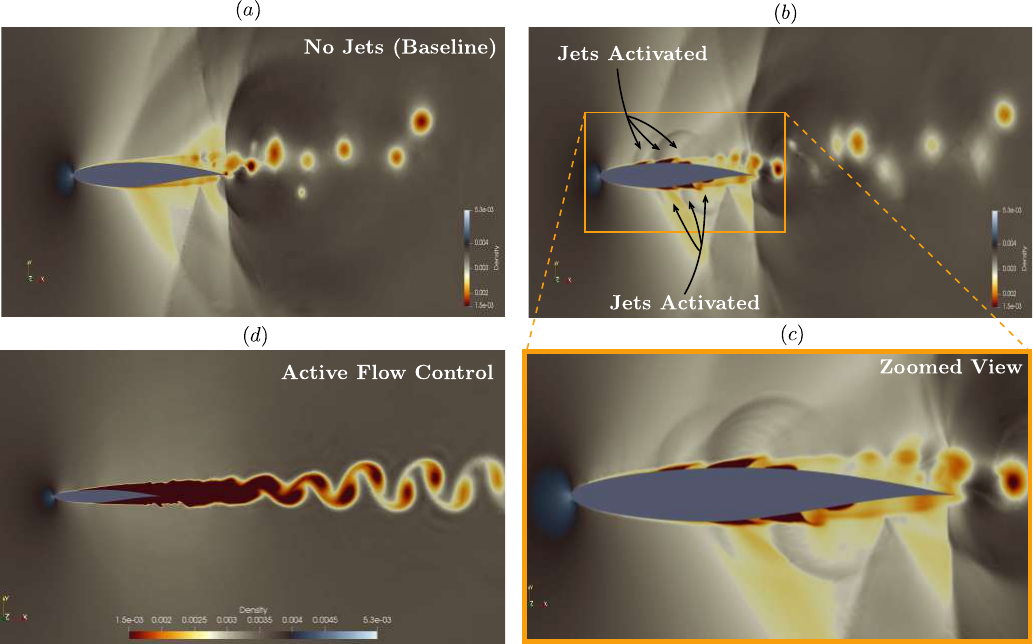}
    \caption{Flow evolution (density contours): $(a)$ Baseline flow without jets, $(b)$ flow with both bottom and top jets activated, $(c)$ zoomed view of the activated jets, and $(d)$ final controlled flow achieving minimum drag.}
    \label{fig:FlowEvo}
\end{figure}

Given that our CFD simulations are computationally demanding, sample efficiency is a critical consideration. Off-policy algorithms are therefore advantageous in this context, as they substantially reduce the number of required environment interactions by reusing prior experiences. In this study, we employ the TD3 algorithm (see Appendix \ref{appD}) in conjunction with a drag reducing and lift-preserving reward function (equation \eqref{RewardE_2}) to examine off-policy learning performance. To mitigate the potential instability inherent in off-policy methods, multiple training runs are conducted under identical conditions to assess the consistency and reliability of the learned policy.

The performance comparison between the on-policy (PPO) and off-policy (TD3) algorithms is summarized in Table \ref{tab:td3_vs_ppo}. For TD3 training, a lift-preserving reward function (Eq. \ref{RewardE_2}) was employed with weighting coefficients $(w_d, w_l) = (2.0, 1.0)$. As shown, the TD3 agent achieves a notable drag reduction of 25.62\%.
Figure \ref{fig:td3_cl_cd} illustrates the temporal evolution of the aerodynamic coefficients during TD3 training. The drag coefficient decreases, converges to a stable value, and keeps maintaining it. Whereas the lift coefficient, though increased significantly by 196.30\% but it exhibits markedly reduced oscillations compared with PPO. Also, the overall lift-to-drag ratio improved by 220.26\%. The increase in lift is reflected in negative reward as shown in Figure \ref{fig:td3_reward}. The evolution of the mass flow rate for all six jets during training is shown in Figure~\ref{fig:td3_mass_flow_rate}. The agent rapidly converges and subsequently maintains a stable mass flow rate with minimal oscillations. Figure \ref{fig:FlowEvo} shows the evolution of the flow (density contours) under jet actuation for drag reduction. Panel (a) depicts the baseline flow without jets, highlighting Kutta waves at the trailing edge and flow separation induced by shock interactions. In panel (b), both bottom and top jets are activated, illustrating their immediate effect on the flow structure. Panel (c) provides a zoomed-in view of the activated jets, revealing local interactions between the injected momentum and the surrounding flow. Panel (d) shows the fully controlled flow, where jet actuation stabilizes the flow, suppresses Kutta-wave formation, and eliminates shock waves, resulting in significant drag reduction. 

 These results indicate that, despite the inherent bias associated with off-policy learning, the TD3 framework, when coupled with a physically informed, lift-preserving reward, achieves a more effective balance between performance and stability. The agent not only attains greater drag reduction but also sustains smoother and more coherent lift dynamics, suggesting enhanced control authority over the unsteady shock–boundary-layer system. This improvement reflects TD3’s ability to exploit richer policy updates and finer exploration in continuous control spaces, enabling the discovery of flow-control strategies that remain robust under the strongly coupled, nonlinear transonic conditions where PPO tends to exhibit oscillatory behavior.

\section{Summary and Discussion}
This study presents the successful application of deep reinforcement learning (DRL) to the active control of transonic flow over the two-dimensional RAE2822 airfoil at $R_e = 50{,}000$. High-fidelity simulations employing a spectral discontinuous Galerkin (DG) method in space and SSPRK(5,4) in time, coupled with adaptive mesh refinement (AMR), accurately capture small-scale flow features such as shock waves, boundary layer separation, and their interactions.  By directly coupling a DRL framework with high-fidelity compressible Navier–Stokes simulations, the agent autonomously developed synthetic-jet control strategies that mitigated shock-induced separation, suppressed unsteady shock oscillations, and improved overall aerodynamic performance. The simulations posed significant computational challenges, including extremely small time steps on the order of $10^{-7}$ due to AMR, which is required to resolve fine flow structures. Rapid actuation frequencies (high $f_a / f_s$) further introduced dynamic mesh adaptation challenges, as shocks and boundary layer separation locations changed abruptly.  Despite these challenges, the DRL agent achieved stable convergence and effective control performance. Importantly, the framework operates without reduced-order modeling or predefined control laws, demonstrating robustness to unsteady flow variations. The \textit{three–three} synthetic-jet configuration, with jets distributed along the upper and lower surfaces of the airfoil, is investigated to effectively control the flow field.
In particular, we employed on-policy \textit{Proximal Policy Optimization (PPO)} algorithm which is one of the popular algorithms employed for active flow control in the literature. First, we designed the reward function to simultaneously minimize the drag coefficient and enhance the lift coefficient. Using this formulation, the DRL agent achieved a 13.78\% reduction in drag and a remarkable 131.73\% increase in lift, resulting in a 121.52\% improvement in the lift-to-drag ratio. Next, the reward function is modified to prioritize drag reduction while maintaining the baseline lift, reflecting typical cruise conditions for passenger aircraft at transonic speeds. Under this scenario, we compared the on-policy PPO algorithm with the off-policy \textit{Twin Delayed Deep Deterministic Policy Gradient (TD3)} algorithm. TD3 outperformed PPO, achieving a 25.62\% reduction in drag. The lift coefficient increased by 196.30\% and exhibited substantially reduced oscillations compared with PPO, resulting in a 220.26\% improvement in the lift-to-drag ratio. These results highlights the inherent challenge of reducing drag without compromising lift in transonic flow, where strong shock–boundary-layer interactions tightly couple drag reduction to variations in pressure distribution and lift.

Future work will focus on enhancing the robustness and generality of the proposed framework. Optimal sensor placement will be investigated in partially observable systems to reflect realistic aircraft conditions and enable efficient practical deployment. The generalization of control policies across flight regimes will be explored by varying angles of attack and Mach numbers, with the aim of developing \textit{global policies} that maintain robust performance under diverse aerodynamic conditions. Extension to three-dimensional wing geometries will allow a more comprehensive evaluation of aerodynamic performance and control complexity.

\appendix
\section{Spectral Discontinous Galerkin Formulation for 2D Compressible Navier-Stokes Equations}\label{appA}
Let the domain be $\Omega = \bigcup_{e=1}^{N_e} \Omega_e$,
and let \( p \ge 0 \) be the polynomial degree. Define the \textit{trial} and \textit{test} (discrete) spaces as the usual discontinuous piecewise polynomial spaces:
\[
\begin{aligned}
\mathcal{V}^h &:= \left\{\, \mathbf{v} \in [L^2(\Omega)]^m \;:\; 
\mathbf{v}|_{\Omega_e} \in [\mathbb{P}^p(\Omega_e)]^m \text{ for each element } \Omega_e \,\right\},\\[6pt]
\mathcal{U}^h &:= \left\{\, \mathbf{w} \in [L^2(\Omega)]^m \;:\;
\mathbf{w}|_{\Omega_e} \in [\mathbb{P}^p(\Omega_e)]^m \text{ for each element } \Omega_e \,\right\}.
\end{aligned}
\]
We take \( \mathcal{U}^h = \mathcal{V}^h \). Here \( m \) is the number of conserved variables, and \( \mathbb{P}^p(\Omega_e) \) denotes the set of all polynomials of total degree \( \le p \) on the element \( \Omega_e \). The approximated solution $\mathbf{U}^h$ can be expressed in terms of the \textit{trial (basis) functions} as
\[
 \mathbf{U}(\mathbf{x},t) \approx \mathbf{U}^h(\mathbf{x},t) = \sum_{i=1}^{N_p} \mathbf{U}_i^h(t)\, \Phi_i^h(\mathbf{x}),
\]
with the corresponding \textit{group flux formulation} as
\[
 \mathbf{F}(\mathbf{x},t) \approx  \mathbf{F}^h(\mathbf{x},t) = \sum_{i=1}^{N_p} \mathbf{F}_i^h(t)\, \Phi_i^h(\mathbf{x}); 
\quad 
 \mathbf{G}(\mathbf{x},t) \approx \mathbf{G}^h(\mathbf{x},t) = \sum_{i=1}^{N_p} \mathbf{G}_i^h(t)\, \Phi_i^h(\mathbf{x}),
\]
where $\Phi_i^h(\mathbf{x})$ are the local basis functions spanning $\mathcal{U}^h|_{\Omega_e}$, and $\mathbf{U}_i^h(t)$, $\mathbf{F}_i^h(t)$, $\mathbf{G}_i^h(t)$ are the corresponding time-dependent coefficients (degrees of freedom) for the solution and fluxes.

Multiplying the conservation form (equation \eqref{NSE}) by a test function $\mathbf{v}^h \in \mathcal{V}^h$ and integrating over the element yields the following weak formulation:
\begin{equation}
\int_{\Omega_e} \mathbf{v}^h \frac{\partial \mathbf{U}^h}{\partial t} \, d\Omega 
+ \int_{\Omega_e} \mathbf{v}^h \left(
\frac{\partial \mathbf{F}_{\text{inv}}^h}{\partial x} 
+ \frac{\partial \mathbf{G}_{\text{inv}}^h}{\partial y}
\right) d\Omega
= \int_{\Omega_e} \mathbf{v}^h \left(
\frac{\partial \mathbf{F}_v^h}{\partial x} 
+ \frac{\partial \mathbf{G}_v^h}{\partial y}
\right) d\Omega.
\end{equation}

Applying integration by parts to the flux terms gives
\begin{multline}
\int_{\Omega_e} \mathbf{v}^h \frac{\partial \mathbf{U}^h}{\partial t} \, d\Omega 
- \int_{\Omega_e} \left(
\frac{\partial \mathbf{v}^h}{\partial x}\mathbf{F}_{\text{inv}}^h 
+ \frac{\partial \mathbf{v}^h}{\partial y}\mathbf{G}_{\text{inv}}^h
\right) d\Omega
+ \int_{\partial\Omega_e} \mathbf{v}^h 
\left(
\mathbf{F}_{\text{inv}}^h n_x 
+ \mathbf{G}_{\text{inv}}^h n_y
\right) dS \\
= -\int_{\Omega_e} \left(
\frac{\partial \mathbf{v}^h}{\partial x}\mathbf{F}_v^h 
+ \frac{\partial \mathbf{v}^h}{\partial y}\mathbf{G}_v^h
\right) d\Omega
+ \int_{\partial\Omega_e} \mathbf{v}^h 
\left(
\mathbf{F}_v^h n_x 
+ \mathbf{G}_v^h n_y
\right) dS.
\end{multline}

Here, $\partial\Omega_e$ denotes the element boundary and $\hat{\mathbf{n}} = (n_x, n_y)^T$ is the outward normal vector.  
Since the solution is discontinuous across element interfaces, the boundary fluxes are replaced by numerical fluxes:
\begin{multline}
\int_{\Omega_e} \mathbf{v}^h \frac{\partial \mathbf{U}^h}{\partial t} \, d\Omega 
= \int_{\Omega_e} \left(
\frac{\partial \mathbf{v}^h}{\partial x}\mathbf{F}_{\text{inv}}^h 
+ \frac{\partial \mathbf{v}^h}{\partial y}\mathbf{G}_{\text{inv}}^h
\right) d\Omega
- \int_{\partial\Omega_e} \mathbf{v}^h 
\left(
\mathbf{F}_{\text{inv}}^{h,*} n_x 
+ \mathbf{G}_{\text{inv}}^{h,*} n_y
\right) dS \\[2mm]
- \int_{\Omega_e} \left(
\frac{\partial \mathbf{v}^h}{\partial x}\mathbf{F}_v^h 
+ \frac{\partial \mathbf{v}^h}{\partial y}\mathbf{G}_v^h
\right) d\Omega
+ \int_{\partial\Omega_e} \mathbf{v}^h 
\left(
\mathbf{F}_v^{h,*} n_x 
+ \mathbf{G}_v^{h,*} n_y
\right) dS.
\end{multline}

The numerical fluxes $\mathbf{F}_{\text{inv}}^{h,*}$, $\mathbf{G}_{\text{inv}}^{h,*}$, $\mathbf{F}_v^{h,*}$, and $\mathbf{G}_v^{h,*}$ couple neighboring elements and account for solution discontinuities at interfaces, computed from the interior state $\mathbf{U}^{h-}$ and exterior state $\mathbf{U}^{h+}$.

The \textit{global system} is obtained by
\begin{align}
\sum_{e=1}^{N_e} 
\Bigg[ 
& \int_{\Omega_e} \mathbf{v}^h \frac{\partial \mathbf{U}^h}{\partial t} \, d\Omega
- \int_{\Omega_e} \Big(
\frac{\partial \mathbf{v}^h}{\partial x} \mathbf{F}_{\text{inv}}^h 
+ \frac{\partial \mathbf{v}^h}{\partial y} \mathbf{G}_{\text{inv}}^h
\Big) d\Omega
+ \int_{\partial\Omega_e} \mathbf{v}^h 
\Big(
\mathbf{F}_{\text{inv}}^{h,*} n_x 
+ \mathbf{G}_{\text{inv}}^{h,*} n_y
\Big) dS \notag\\
& + \int_{\Omega_e} \Big(
\frac{\partial \mathbf{v}^h}{\partial x} \mathbf{F}_v^h 
+ \frac{\partial \mathbf{v}^h}{\partial y} \mathbf{G}_v^h
\Big) d\Omega
- \int_{\partial\Omega_e} \mathbf{v}^h 
\Big(
\mathbf{F}_v^{h,*} n_x 
+ \mathbf{G}_v^{h,*} n_y
\Big) dS
\Bigg] = 0.
\end{align}

\subsection{Flux Reconstruction at Element Interfaces }
The DG weak formulation requires numerical fluxes to handle discontinuities across elements and to evaluate boundary integrals. 

\vspace{0.2cm}
\noindent\textbf{Inviscid Flux:} At element interfaces, solution jumps are handled using the Lax-Friedrichs (Rusanov) flux (\cite{edwards2006dominant}):
\[
\mathbf{F}_{\text{inv}}^{h,*} = \frac{1}{2} \left[ \mathbf{F}^h(\mathbf{U}^{h-}) + \mathbf{F}^h(\mathbf{U}^{h+}) \right] - \frac{\lambda}{2} (\mathbf{U}^{h+} - \mathbf{U}^{h-}),
\]
where $\mathbf{U}^{h-}$ and $\mathbf{U}^{h+}$ are the interior and exterior states, respectively, and $\lambda$ is the maximum wave speed at the interface. The first term averages the fluxes, while the second adds dissipation proportional to the solution jump, stabilizing the scheme near discontinuities. Similarly, $\mathbf{G}_{\text{inv}}^{h,*}$ can be calculated. 

\vspace{0.2cm}
\noindent\textbf{Viscous Flux:} At element interfaces, the numerical viscous fluxes $\mathbf{F}_v^{h,*}$ and $\mathbf{G}_v^{h,*}$ are computed using the BR1 scheme (\cite{bassi1997high}).
To evaluate the viscous fluxes in the DG framework, an auxiliary gradient variable 
\(\mathbf{S}^h = \nabla \mathbf{U}^h\) is introduced. This allows the viscous flux at an element interface to be expressed as a function of both the solution and its gradient:
\[
\mathbf{F}_v^{h,*} = \frac{1}{2} \left[ \mathbf{F}_v(\mathbf{U}^{h-}, \mathbf{S}^{h-}) + \mathbf{F}_v(\mathbf{U}^{h+}, \mathbf{S}^{h+}) \right]. 
\]
Similarly, $\mathbf{G}_v^{h,*}$ can be calculated.

The auxiliary gradient \(\mathbf{S}^h\) is computed locally for each element using the DG gradient operator:
\[
\int_{\Omega_e} \phi_j \, \mathbf{S}^h \, d\Omega 
= - \int_{\Omega_e} \nabla \phi_j \, \mathbf{U}^h \, d\Omega 
+ \int_{\partial \Omega_e} \phi_j \, \hat{\mathbf{U}} \, \hat{\mathbf{n}} \, dS,
\]
where \(\phi_j\) are the test functions and \(\hat{\mathbf{U}}\) is the numerical trace of \(\mathbf{U}^h\) on the element boundary, typically taken as the interior value or an average with neighboring elements.  
This formulation ensures that the computed gradient is consistent with the DG framework, allowing the viscous fluxes to account for solution gradients while maintaining local conservation and stability across non-conforming interfaces.

Spectral DG formulation is achieved by employing high-order polynomial basis functions whose interpolation and flux evaluations are collocated at Gauss–Lobatto–Legendre (GLL) nodes within each element. By using Lagrange polynomials through these nodes and computing the weak form integrals with Gauss–Lobatto quadrature (\cite{hesthaven2008nodal}), the method ensures exact integration at the collocation points. This approach preserves the local conservation and flexibility of standard DG while enabling spectral (exponentially fast) convergence for smooth solutions, effectively combining the strengths of spectral and discontinuous Galerkin methods.

\section{Time-discretization using SSPRK Method}\label{appB}
The semi-discrete DG formulation results in a system of ordinary differential equations (ODEs) in time:
\begin{equation}
\frac{d\mathbf{U}^h}{dt} = \mathcal{L}(\mathbf{U}^h, t)
\end{equation}
where $\mathbf{U}^h$ is the vector of discrete solution values and $\mathcal{L}$ represents the spatial discretization operator including the volume and surface flux terms. The general form of an $s$-stage SSPRK method (\cite{spiteri2002new}) applied to the semi-discrete DG system is:
\begin{align}
\mathbf{U}^{h,(0)} &= \mathbf{U}^{h,n}, \\
\mathbf{U}^{h,(i)} &= \sum_{k=0}^{i-1} \left[\alpha_{ik} \mathbf{U}^{h,(k)} + \Delta t \, \beta_{ik} \, \mathcal{L}(\mathbf{U}^{h,(k)})\right], \quad i = 1, 2, \ldots, s, \\
\mathbf{U}^{h,n+1} &= \mathbf{U}^{h,(s)},
\end{align}
where $\mathbf{U}^{h,(i)}$ denotes the discrete solution at intermediate stage $i$, and the coefficients $\alpha_{ik}$ and $\beta_{ik}$ satisfy $\alpha_{ik} \geq 0$, $\beta_{ik} \geq 0$, and $\sum_{k=0}^{i-1} \alpha_{ik} = 1$ for consistency. 

The CFL (Courant-Friedrichs-Lewy) number for the SSPRK54 scheme is 
$\tilde{c} = \min_{i,k} \frac{\alpha_{ik}}{\beta_{ik}} \approx 1.508,$
which sets the maximum allowable time step:
\begin{equation}
\Delta t_C \leq \tilde{c} \, \frac{\Delta x_{min}}{(2p+1)\lambda_{\max}},
\end{equation}
where $\Delta x$ is the characteristic element size and $\lambda_{\max}$ is the maximum wave speed in the domain (as computed in the surface flux section). 
For diffusion/viscous term
\begin{equation}
\Delta t_D \leq \tilde{d} \, \frac{\Delta x^2_{min}}{(2p+1)^2\nu_{\max}},
\end{equation}
with $\tilde{d} \approx 0.5$ and $\nu_{\max}$ is the larger of kinematic viscosity and thermal diffusivity.

The time step, $\Delta t$, is determined as the minimum of the convective (hyperbolic) and diffusive (parabolic) CFL constraints: 
$\Delta t = \text{min} (\Delta t_C,\Delta t_D) $.
SSPRK54 provides fourth-order temporal accuracy while preserving strong stability, essential for resolving shocks and high-gradient features, such as
discontinuities in compressible transonic flows.

\section{Airfoil Jet Details}\label{appC}
\begin{figure}[htpb]
     \centering \includegraphics[scale=0.8, clip=true]{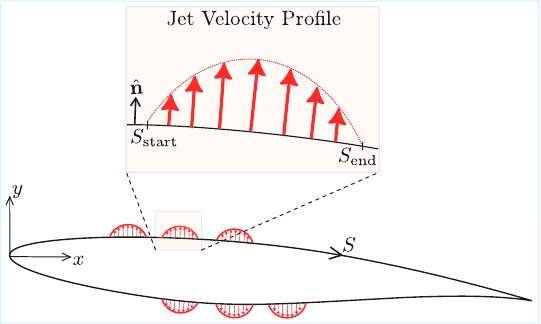}
     \caption{Velocity profile of jets along the airfoil surface.}\label{fig:jet_profile}
 \end{figure}

The jet velocity distribution follows a sinusoidal profile across each jet width to ensure smooth flow transitions at the jet boundaries; see figure \ref{fig:jet_profile}. Let \( S \) denote the coordinate measured along the airfoil surface (chord-wise distance following the surface contour). 
For a jet spanning from \( S_{\text{start}} \) to \( S_{\text{end}} \), the local jet velocity magnitude is defined as:
\begin{equation}
\mathbf{u}_{\text{jet}}(S) = \mathbf{u}_{\text{max}} \cdot \sin\left(\pi \frac{S - S_{\text{start}}}{S_{\text{end}} - S_{\text{start}}}\right),
\end{equation}
where \( \mathbf{u}_{\text{max}} = \lambda_{u} \, \mathbf{u}_{\infty} \) is the maximum jet velocity magnitude at the jet center,  
\( \lambda_{u} \in [-1, 1] \) is the velocity factor controlled by the RL agent, and \( \mathbf{u}_{\infty} =\{u_{\infty},v_{\infty}\} \) is the freestream velocity magnitude.
The figure \ref{fig:jet_profile} illustrates the velocity profile of each jet. The corresponding local jet velocity vector is expressed as:
\begin{equation}
\mathbf{u}_{\text{jet,local}}(S) = \mathbf{u}_{\text{jet}}(S) \cdot \hat{\mathbf{n}} = 
\begin{bmatrix} 
u_{\text{jet,local}}^x(S) \\
u_{\text{jet,local}}^y(S)
\end{bmatrix},
\end{equation}
where \( \hat{\mathbf{n}} \) denotes the local jet direction at the airfoil surface. Each local jet velocity $\mathbf{u}_{\mathrm{jet,local}}$ is rotated into the global frame to account for the airfoil’s angle of attack $\alpha$:
\[
\begin{bmatrix} u_{\mathrm{jet}}^x \\[1ex] u_{\mathrm{jet}}^y \end{bmatrix} = 
\begin{bmatrix} 
\cos\alpha & -\sin\alpha \\[0.5ex]
\sin\alpha & \cos\alpha
\end{bmatrix}
\begin{bmatrix} u_{\mathrm{jet,local}}^x \\[1ex] u_{\mathrm{jet,local}}^y \end{bmatrix}.
\]  
The resulting $u_{\mathrm{jet}}^x$ and $u_{\mathrm{jet}}^y$ values modify the boundary conditions to incorporate jet effects in the flow.

Jet angles are fixed during training to ensure stable learning, with top- and bottom-surface jets set perpendicular to the local airfoil surface ($90^\circ$ and $-90^\circ$, respectively). To maintain \textit{zero net mass flux} through the airfoil, the middle jets (Jets~2 and~5) act as compensators, with their velocities determined by mass conservation: 
$\sum_{i=1}^3 Q_i = 0 \quad \text{(top surface)}, $ and $ 
\sum_{i=4}^6 Q_i = 0 \quad \text{(bottom surface)},$
where $Q = \rho_{\infty} \mathbf{u}_{\text{jet,local}} \cdot \mathbf{A}_{\text{jet}}$ is the mass flow rate, $ \mathbf{A}_{\text{jet}}$ is the area vector of the jet cross-section, and the index $i$ denotes the $i^{\text{th}}$ jet.

\section{Twin Delayed Deep Deterministic Policy Gradient (TD3)}\label{appD}

The TD3 algorithm~\cite{fujimoto2018addressing} is an off-policy actor–critic method designed for continuous control tasks. It extends the DDPG framework~\cite{lillicrap2015continuous} through three key modifications aimed at mitigating overestimation bias and enhancing training stability:

\begin{itemize}
    \item \textbf{Clipped double Q-learning:} employs two critic networks and computes the target value using the minimum of their Q-estimates, reducing overoptimistic bias in value updates.
    \item \textbf{Delayed policy updates:} updates the actor network less frequently than the critics, improving the stability of policy learning.
    \item \textbf{Target policy smoothing:} adds small, clipped noise to the target actions during critic updates, reducing sensitivity to value spikes and promoting smoother value estimation.
\end{itemize}

\begin{algorithm}[H]
\caption{TD3 for Continuous Learning}
\begin{algorithmic}[1]
\STATE Initialize actor network $\pi_\theta$, critics $Q_{\phi_1}, Q_{\phi_2}$ and target networks
\STATE Initialize replay buffer $\mathcal{D}$
\WHILE{training not stopped}
    \STATE Observe current state $s_t$
    \STATE Select action $a_t = \pi_\theta(s_t) + \epsilon$, \text{with } $\epsilon \sim \text{Normal}(0, \sigma)$
    \STATE Apply $a_t$ to CFD simulation, get next state $s_{t+1}$ and reward $r_t$
    \STATE Store transition $(s_t, a_t, r_t, s_{t+1})$ in buffer $\mathcal{D}$
    \STATE Sample minibatch of transitions from $\mathcal{D}$
    \STATE Compute target actions:
        \STATE \quad $\tilde{a}_{t+1} = \pi_{\theta'}(s_{t+1}) + \text{clip}(\text{Normal}(0, \sigma_{\text{target}}), -c, c)$
    \STATE Compute target value:
        \STATE \quad $y_t = r_t + \gamma \min(Q_{\phi_1'}(s_{t+1}, \tilde{a}_{t+1}), Q_{\phi_2'}(s_{t+1}, \tilde{a}_{t+1}))$
    \STATE Update critics $Q_{\phi_i}$ by minimizing MSE loss
    \IF{current training step \% policy\_update\_interval == 0}
        \STATE Update actor (policy network) using the critic's gradients
        \STATE Update target networks gradually (soft update):
        \STATE \quad For each critic: $\phi_i' \gets \tau \phi_i + (1-\tau) \phi_i'$
        \STATE \quad For actor: $\theta' \gets \tau \theta + (1-\tau) \theta'$
    \ENDIF
\ENDWHILE
\end{algorithmic}
\end{algorithm}

In summary, TD3 enhances sample efficiency and stability in continuous control by explicitly addressing the overestimation bias inherent in DDPG. Nevertheless, as an off-policy algorithm, it remains susceptible to bias arising from discrepancies between replay buffer data and the current policy, which can affect convergence. Multiple training runs are therefore recommended to ensure consistent and reliable policy performance.


\bibliographystyle{jfm}
\bibliography{jfm}


\end{document}